\documentclass[prd,twocolumn,showpacs,groupedaddress,superscriptaddress,nofootinbib,floatfix,preprintnumbers,longbibliography]{revtex4-1}

\usepackage{amssymb,amsmath,graphicx,color,amsfonts}
\usepackage{bm}
\usepackage[tight]{subfigure}
\usepackage[export]{adjustbox}
\usepackage{braket}
\usepackage[colorlinks=true,citecolor=blue,linkcolor=blue]{hyperref}
\usepackage[table]{xcolor}
\usepackage{cancel}
\usepackage{multirow}
\usepackage{fixltx2e}
\usepackage{blindtext}

\newcounter{oldtocdepth}

\newcommand\beq{\begin{equation}}
\newcommand\eeq{\end{equation}}
\newcommand\bea{\begin{eqnarray}}
\newcommand\eea{\end{eqnarray}}

\begin{document}
\title{
Cold Atom Quantum Simulator for String and Hadron Dynamics in Non-Abelian Lattice Gauge Theory}

\author{Raka Dasgupta}
\affiliation{Dept. of Physics, University of Calcutta, 92 A. P. C. Road, Kolkata-700009, India}
\email{dasguptaraka@gmail.com}

\author{Indrakshi Raychowdhury}
\affiliation{
Maryland Center for Fundamental Physics and Department of Physics, 
University of Maryland, College Park, MD 20742, USA}
\email{iraychow@umd.edu}

\date{\today}

\preprint{UMD-PP-020-8}

\begin{abstract}
We propose an analog quantum simulator for simulating real time dynamics of $(1+1)$-d non-Abelian gauge theory well within the existing  capacity of ultracold atom experiments. The scheme calls for the realization of a two-state ultracold fermionic system in a 1-dimensional bipartite lattice, and the observation of subsequent tunneling dynamics.  Being based on novel loop string hadron formalism of SU(2) lattice gauge theory, this  simulation technique is completely SU(2) invariant and simulates accurate dynamics of physical phenomena such as string breaking and/or pair production. The scheme is scalable, and particularly effective in simulating the theory in weak coupling regime, and also bulk limit of the theory in strong coupling regime up to certain approximations.
This paper also presents a numerical benchmark comparison of exact spectrum and real time dynamics of lattice gauge theory to that of the atomic Hamiltonian with experimentally realizable range of parameters. 
\end{abstract}
\maketitle

\section{Introduction}
\label{sec:intro}
\noindent


Gauge field theories constitute an exceptionally  powerful theoretical framework that describes three of the four fundamental interactions of nature. Non-Abelian gauge symmetry lies at the heart of standard model of particle physics. Quantum chromodynamics (QCD) is a SU(3) gauge theory that accurately describes interactions between quarks via gluons as mediators. However, the phenomenon of quark confinement is yet to be established analytically  from the fundamental theories in physics. In 1974, Wilson proposed a regularization of the gauge theory on space-time lattices \cite{Wilson:1974sk} that exhibits the phenomenon of confinement in the strong coupling limit. Wilson's lattice gauge theory (LGT) is an extremely successful non-perturbative technique that has been used extensively over past four-five decades. One of the reasons behind this versatility and success story of LGT is that one can perform lattice QCD calculations by Monte Carlo simulations
\cite{CREUTZ1983201} and world's largest super-computing resources are now being employed for the same \cite{joo2019status}.

Although  the lattice QCD program based on numerical calculations is really successful, there is the infamous `sign problem' that forbids lattice QCD calculations \cite{de2010simulating} to work at certain  regimes such as to address systems with finite and non-zero density or calculating real time dynamics. The natural framework to study dynamics of gauge theory is Hamiltonian framework. However, the exponential growth of Hilbert space dimension with system size for any quantum system claims that classical computer may not be the best computational tool \cite{wiese2013ultracold} for Hamiltonian simulation.  Feynman's visionary idea \cite{feynman1982simulating} combined with  recent technological advancement in quantum information science and technology leads to a perfect alternative in the form of  quantum computation and/or simulation, and offers hope for settling these issues. 

The concept of analog quantum simulation involves mimicking a quantum system described by a Hamiltonian (simulated Hamiltonian) by another completely different quantum system described by some other Hamiltonian (simulating Hamiltonian). The idea is, if it is hard to analyze the first Hamiltonian  mathematically or numerically, it can be mapped to the second Hamiltonian, and that Hamiltonian can be studied in an experiment by suitably tuning the parameters. Systems of ultracold atoms \cite{bloch2012quantum}  or ions  \cite{blatt2012quantum}  trapped in optical lattices serve as excellent quantum simulators, as the relevant parameters can be precisely measured and controlled. In the past few years, there has been a steady progress in this direction in  the context of gauge theory.  \cite{zohar2011confinement,Zohar:2012xf, Zohar:2013zla,Zohar:2015hwa,Banerjee:2012pg,Banerjee:2012xg,stannigel2014constrained,Gonzalez-Cuadra:2017lvz,tagliacozzo2013simulation, kasper2016schwinger,schweizer2019floquet,mil2020scalable,yang2020observation,davoudi2020towards}. 

 The experimental realization of Bose-Einstein condensate in 1995 \cite{anderson1995observation, bradley1995evidence, davis1995bose} and the  subsequent exploration of trapped  ultracold fermions \cite{demarco1999onset,schreck2001quasipure, truscott2001observation, o2002observation}  in the degenerate regime paved the way towards newer directions of atomic and molecular physics. In the initial years, theoretical and experimental works in this field mostly focused on the emergence of macroscopic quantum coherent phenomena in a many-body system, and the formation of molecular condensates \cite{greiner2003emergence,jochim2003bose,zwierlein2003spectroscopic} . However, it was soon discovered that  ultracold atomic systems can serve as wonderful testing grounds for other branches of physics as well, by virtue of the sheer tunability of the parameters. The atom-atom scattering length (and thus, the interaction strength ) in ultracold gases can be varied across a wide range  by methods of Feshbach resonances. The creation of optical lattices \cite{grein} by using two counter-propagating coherent laser beams took this tunability a step further : as the size, shape and dimensionality of the lattice could be easily controlled. Advanced cooling and trapping methods has now led to  temperatures as low as in nanoKelvin and picoKelvin ranges, resulting in quantum engineering at an unparalleled precision level. This allows for  each individual atom to be  monitored, and one can have a perfect quantum simulator.  In the past, cold atomic systems have successfully emulated a rich variety of systems and addressed problems in disordered systems, spin liquids, superconductivity, nuclear pairing, artificial gauge fields and topology \cite{lewenstein2007ultracold,lewenstein2012ultracold,gross2017quantum}.

Over the last few years, there has been a continuous pursuit  towards analog quantum simulating lattice gauge theories using cold atom systems.  During the first half of the past decade, these proposals mostly focused on  constructing quantum simulators for both Abelian and non-Abelian gauge theories in Kogut-Susskind formalism \cite{zohar2011confinement, Zohar:2012xf, Zohar:2013zla, Zohar:2015hwa}  as well as Quantum Link Model formulation \cite{Banerjee:2012pg, Banerjee:2012xg, stannigel2014constrained} and also Abelian Higgs Model in 2+1 dimensions \cite{Gonzalez-Cuadra:2017lvz}. All these schemes involved a careful designing of the set-up so that the system  remains in the gauge invariant Hilbert space throughout the dynamics. There was also one generic experimental proposal for quantum simulating non-Abelian gauge theories using Rydberg atom gates \cite{tagliacozzo2013simulation}. A general feature of all of these proposals are : the lattice sites (for matter field) and the links (for gauge field) of the original lattice gauge theory are simulated by   bosonic and/or fermionic atoms trapped in arrays of potential wells of optical lattices; there being a one-to-one correspondence between the LGT sites and the sites of the spatial optical lattice. Additionally, auxiliary atoms were also considered to effectively create the plaquette term of the gauge theory Hamiltonian \cite{Zohar:2012xf, Zohar:2015hwa,Gonzalez-Cuadra:2017lvz}. However, none of these proposals has yet been experimentally implemented or  numerically benchmarked. We note that the proposals involving quantum link models are more suited for practical realization. The amount of information of QED actually captured in finite dimensional Hilbert space of Quantum Link Model and its real time dynamics  was studied in \cite{kasper2016schwinger}. 

Following the first experimental demonstration of a digital quantum simulation of a lattice  Schwinger model \cite{martinez2016real},  the first analog quantum simulation of $\mathbb Z_2$ gauge theory on a two staggered site lattice by cold atom quantum simulator was reported in \cite{schweizer2019floquet}. However, generalization of this scheme either to make it scalable or to simulate theory with continuous gauge groups has not yet been reported. The first experiment demonstrating a scalable quantum simulation of continuous gauge theory was reported very recently \cite{mil2020scalable} that demonstrates the engineering of an elementary building block of the U(1) quantum link model in 1 spatial dimension using a mixture of bosonic atoms.  This scheme is quite unique,  as unlike the past proposals, a single potential well here does not merely hosts a site or link of gauge theory. Instead, a physical site here contains  both the matter and the gauge states in the form of two different atoms, and also the matter-gauge field interactions : thus acting as the fundamental building block of the gauge theory Hamiltonian. Exact Implementation of gauge invariance is another challenging task in any of these quantum simulators.  Recently U(1) gauge invariance in the simulated system has been observed experimentally for a long lattice \cite{yang2020observation}.  In spite of these significant advances, a practically realizable analog quantum simulation scheme for simulating dynamics of a gauge theory even with the simplest non-Abelian, continuous gauge group such as SU(2) is absent in past literature. 

In this paper, we present a scalable and immediately realizable quantum simulation proposal for quantum simulating SU(2) gauge theory coupled to fermionic staggered matter in $(1+1)$-d. This work also provides numerical study of the spectrum of the simulating and simulated Hamiltonian and compare their real time dynamics for a small lattice.

We aim to quantum simulate the Kogut-Susskind (KS) Hamiltonian for LGT \cite{Kogut:1974ag}. In a recent study \cite{new}, it has been demonstrated that amongst many variants of Hamiltonian formulation of non-Abelian gauge theories \cite{chandrasekharan1997quantum,Brower:1997ha,Zohar:2019ygc,Zohar:2018cwb, Zohar:2014qma,banuls2017efficient},  the loop string hadron (LSH) formalism \cite{Raychowdhury:2019iki}  is the most convenient and computationally least expensive one for $(1+1)$-d within the scope of classical computation. The reason is, being a fully gauge invariant formalism, the LSH Hamiltonian describes the dynamics of only relevant physical degrees of freedom. In 1d spatial lattice, that is precisely the dynamics of strings and hadrons.
It can be shown \cite{new, sala2018variational} that  in 1+1d, any gauge theory with open boundary condition can be mapped to a theory of only fermions, i.e., equivalent to the XYZ model and hence much simpler to analyze. The novel LSH formalism shares many features of this purely fermionic formalism but can  actually be generalized to periodic boundary conditions as well as to higher dimensions \cite{Raychowdhury:2019iki}. 

The present paper exploits this versatility of  LSH formalism of SU(2) gauge theory. Here, different parameter regimes of SU(2) gauge theory  are  mapped to different parameter regimes of an atomic Hamiltonian: that of an ionic Hubbard model with different total number of fermions on the lattice. We consider the half filled Hubbard model that is exactly equivalent to the gauge theory Hilbert space containing strong coupling vacuum. We show that the spectrum, obtained with exact diagonalization of both the simulating and simulated system compared remarkably in the weak coupling regime.   We also provide a benchmark comparison of the dynamics of atomic system directly mapped to the pair production-string breaking dynamics of the low energy sector of SU(2) gauge theory. The numerical analysis employs parameters and experimental set-ups already realized with ultracold atom systems. We demonstrate two key points:  i) the full gauge theory Hamiltonian can be reduced to an approximated LSH Hamiltonian, which, in turn, can be perfectly mimicked by the atomic system to the low energy dynamics in the weak coupling limit of gauge theory, and ii) for the strong and intermediate coupling regimes,  the difference between the full gauge theory Hamiltonian and the approximated Hamiltonian is slightly more prominent. But one can still access dynamics of strings and hadrons in presence of a background gauge field in the bulk limit of the lattice  by tuning the on-site interaction parameter of the Hubbard Hamiltonian. Further improvements of this scheme to include dynamical gauge fields in higher dimension, and also generalization to SU(3) gauge theory will take us close to quantum simulating the full QCD.


In this proposal, a system of ultracold fermions trapped in optical lattices is considered as quantum simulating platform for non-Abelian gauge theories. The plan of the paper is as follows: In section \ref{sec:KS1} we briefly discuss lattice gauge theory Hamiltonian including LSH framework in general and also in the weak coupling limit. In section \ref{sec:atomic} we discuss the atomic system to be used for the quantum simulation scheme, a fermionic Hubbard model on a bipartite lattice. In section \ref{sec:parameter} we map the gauge theory Hamiltonian to the Hubbard model Hamiltonian introduced before for both weak and strong coupling regimes of gauge theory. In section \ref{sec:exp}  the proposed experimental set-up is described. Section \ref{sec:dynamics} contains numerical study and comparison of the spectrum  and real time dynamics of both the simulating and simulated systems using the parameters for the proposed experimental scheme. Finally, in section \ref{sec:conclusion} we discuss our results and also future prospects.


\section{SU(2) lattice gauge theory in $(1+1)$-d}
\label{sec:KS1}
\noindent
 Hamiltonian or canonical formulation of lattice gauge theories was developed by Kogut and Susskind \cite{Kogut:1974ag} right after Wilson introduced lattice gauge theory originally in Euclidean formalism \cite{Wilson:1974sk}. While, classical computing for lattice gauge theory has explored the original Euclidean formulation, the Hamiltonian framework, being not much useful in classical computation era  remain unexplored. However, the interest in exploring Hamiltonian description of lattice gauge theories is renewed, as it turns out to be  the natural framework to work with in the upcoming quantum simulation/computation era. The mostly used formalism in this context is Quantum Link Model representation of gauge theory as it provides a finite dimensional representation of the gauge fields. There is a drawback though :  in smaller dimensions, that are accessible by present-day  quantum technology, the quantum link model does not have the desired spectrum as obtained with the original Kogut-Susskind Hamiltonian \cite{chandrasekharan1997quantum,new}. In this work we consider the original Kogut-Susskind Hamiltonian for simplest non-Abelian gauge group, i.e. SU(2) and proceed to construct a quantum simulator for the same in $(1+1)$-d.

The Kogut-Susskind (KS) Hamiltonian describing SU(2) Yang Mills theory coupled to staggered fermions on $(1+1)$-d (1d spatial lattice and continuous time) \cite{Kogut:1974ag} can be written as:
\begin{eqnarray}
H^{({\rm KS})}&=&H^{({\rm KS})}_E+H^{({\rm KS})}_M+H^{({\rm KS})}_I.
\label{eq:HKS}
\end{eqnarray}
Where, $H^{({\rm KS})}_E$ corresponds to electric part of the Hamiltonian given by,
\begin{eqnarray}
H^{({\rm KS})}_E&=& \frac{g^2a}{2} \sum_{j=0}^{N-1} \sum_{a=1}^3{E^a}(j){E^a}(j).
\label{eq:HEKS}
\end{eqnarray}
Here, $$\sum_{a=1}^3{E^a}(j){E^a}(j)=\sum_{a=1}^3{E_L^a}(j){E_L^a}(j)=\sum_{a=1}^3{E_R^a}(j){E_R^a}(j)$$ for left and right electric fields $\mathbf{E}_{L/R}$ associated with a link connecting sites $j$ and $j+1$.

The staggered fermionic matter $\psi$, in the fundamental representation of $SU(2)$ consisting of two components  $\bigl( \begin{smallmatrix}\psi_1\\ \psi_2\end{smallmatrix}\bigr)$ yields a staggered mass term:
\begin{eqnarray}
H^{({\rm KS})}_M&=&m\sum_{j=0}^{N}(-1)^j\left[\psi^{\dagger}(j)\cdot \psi(j) \right].\label{eq:HMKS}
\end{eqnarray}
$H^{({\rm KS})}_I$ denotes interaction between the fermionic and gauge fields and is given by:
\begin{eqnarray}
H^{({\rm KS})}_I&=&\frac{1}{2a}\sum_{j=0}^{N-1}\left[\psi^{\dagger}(j) U(j) \psi(j+1)+{\rm h.c.} \right]
.\label{eq:HIKS}
\end{eqnarray}
 The gauge link $U(j)$ is a $2\times2$ unitary matrix defined on the link connecting  sites $j$  and $j+1$. A temporal gauge is chosen to derive the above Hamiltonian which sets the gauge link along the temporal direction equal to unity.

The color electric fields $E_{L/R}^{a}$ are defined at the left $L$ and right $R$ sides of each link and they satisfy the following commutation relations (su(2) algebra) at each end:
\begin{eqnarray}
[E_L^a(j),E_L^b(j')]&=&i\epsilon^{abc}\delta_{jj'} E_L^c(j),
\nonumber\\
{[E_R^a(j),E_R^b(j')]}&=&i\epsilon^{abc} \delta_{jj'} E_R^c(j'),
\nonumber\\
{[E_L^a(j),E_R^b(j')]}&=&0,
\label{eq:ERELcomm}
\end{eqnarray}
where $\epsilon^{abc}$ is the Levi-Civita symbol.  The electric fields and the gauge link satisfy the following quantization conditions at each site,
\begin{eqnarray}
[E_L^a(j),U(j')]=-\frac{\sigma^a}{2}\delta_{jj'}U(j),
\nonumber\\
{[E_R^a(j),U(j')]}= U(j)\frac{\sigma^a}{2},
\label{eq:EUcomm}
\end{eqnarray}
where $\sigma^a$ are the Pauli matrices. 
The Hamiltonian in (\ref{eq:HKS}) is gauge invariant as it commutes with the  Gauss' law operator,
\begin{equation}
G^a(j)=E^a_L(j)+E^a_R(j-1)+\psi^\dagger(j) \frac{\sigma^a}{2} \psi(j)
\label{eq:Ga}
\end{equation}
at each site $j$.  The physical sector of the Hilbert space corresponds to the space consisting of states annihilated by (\ref{eq:Ga}).

For LGT, the natural and most convenient basis is formed out of eigenstates of the electric-field operator. Tensor product of fermionic occupation number basis and electric field basis constitute the full Hilbert space. This particular basis, being eigenbasis of the diagonal Hamiltonian ($H_E+H_M$)  in the $g \to \infty$ limit, is called the strong-coupling basis of LGT. 

In the strong coupling regime, lattice gauge theory shows desired physics such as quark confinement and finite mass gap. In this limit, along with finite lattice spacing $a$, the interaction terms in (\ref{eq:HIKS}) that involves transitions between different eigenstates of the electric-field operator becomes insignificant, and hence in the Hamiltonian, diagonal terms dominate over the off-diagonal ones in the strong coupling basis. As $g\to \infty$, only very small electric flux configurations on the lattice contribute to the low energy sector of the theory. In this regime, lattice Hamiltonian matrices can be analyzed perturbatively with the electric  part as the unperturbed Hamiltonian. Order by order perturbation corrections yield a  finite dimensional Hilbert space, within a cut-off imposed on the bosonic quantum number corresponding to gauge flux.  The computation cost rises exponentially with increasing Hilbert space dimension, that grows with system size as well as cut-off \cite{new}. As a result, calculating Hamiltonian dynamics for an arbitrary large system even with the largest possible computer seems impossible. 
However, the continuum limit of the LGT lies in the opposite regime, where $g\rightarrow 0, a\rightarrow 0$ together with bulk limit , i.e lattice size $N\rightarrow \infty$. In this regime, the dynamics becomes too much cut-off sensitive, all possible electric flux states do contribute to the low energy spectrum of the theory with major contribution coming from strong coupling states with electric flux values to be larger and larger with $g\to 0$. The Hamiltonian moves away from diagonal structure as (\ref{eq:HIKS}) becomes dominant with $a\to 0$. As a whole, analyzing the weak coupling limit of lattice gauge theory is extremely difficult on a classical computer except some extrapolation technique of strong coupling analysis. 


In a very recent work \cite{new}, all available formalisms for non-Abelian gauge theory with gauge group SU(2) in $(1+1)$-d has been analyzed and compared in terms of their applicability in Hamiltonian simulation. As concluded in \cite{new}, the recently developed LSH formalism \cite{Raychowdhury:2019iki} enjoys two unique advantages: i) It is exactly equivalent to the original Kogut-Susskind Hamiltonian ii) it removes the non-trivial steps (computational costs) required in the original Hamiltonian formulation to contain the dynamics in the gauge invariant sector of LGT Hilbert space. The second advantage becomes particularly important in designing analog/digital quantum simulator \cite{Raychowdhury:2018osk, Yang:2020yer}. That is why we choose  the novel LSH framework to describe gauge theory and map the same to an atomic Hamiltonian. It is already established \cite{new} that the original Kogut-Susskind Hamiltonian described here, and the LSH Hamiltonian (to be described in next section) share identical spectrum and hence generate the same dynamics.  At this point we must mention that all of the feasible/implemented previous proposals involve QLM formulation of lattice gauge theory, that in lower dimension exhibits a complete different spectrum as well as a different Hilbert space than that of the Kogut-Susskind Hamiltonian.

\subsection{Loop-String-Hadron (LSH) Hamiltonian}
\label{sec:KS}
\noindent
LSH formalism of lattice gauge theory is based on prepotential framework, where, the original canonical conjugate variables of the theory, i.e color electric field and link operators are replaced by a set of harmonic oscillator doublets, defined at each end of a link \cite{Mathur:2004kr,Mathur:2007nu,Mathur:2010wc,Anishetty:2009ai,Anishetty:2009nh,Anishetty:2014tta,Raychowdhury:2013rwa,Raychowdhury:2014eta,Raychowdhury:2018tfj}. In prepotential framework, the SU(2) gauge group is  confined to each lattice site allowing one to have local gauge invariant operators and states  at each site. For pure gauge theory, these local gauge invariant operators and states can be interpreted as local snapshots of Wilson loop operators of original gauge theory. One can now construct local loop Hilbert space by action of local loop operators on strong coupling vacuum of the theory (no flux state) defined locally at each site. At this point, we must mention that, mapping the local loop picture to original loop description of gauge theory requires one extra constraint on each link, that states
\bea
\label{AGL}
N_L(j)=N_R(j)
\eea
where, $N_{L(R)}$ is occupation number of prepotentials/Schwinger bosons at the left(right) end of a link connecting sites $j$ and $j+1$. This constraint is actually a consequence of the constraint ${\bm E}_L^2=\bm{E}_R^2$ mentioned in section \ref{sec:KS1}.          

Inclusion of staggered fermionic matter fields for SU(2) gauge theory at each lattice site, combines smoothly with local loop description obtained in prepotential framework as both the prepotential Schwinger bosons and matter fields transform as fundamental representation of the local SU(2) at that site. In addition to local gauge invariant loop operators, one can now combine matter and prepotentials to construct local string operators, that denotes start of a string from a particle and/or end of a string at an antiparticle. Matter fields combine into local gauge invariant configurations representing hadrons likewise in the original formalism. This complete description is named as LSH formalism as in \cite{Raychowdhury:2019iki}. We are not going into the details of the full LSH formalism here. Instead, we will focus on the application of LSH formulation to one spatial dimension only, and describe the appropriate framework.

\begin{figure}[t]
\includegraphics[width=0.485\textwidth]{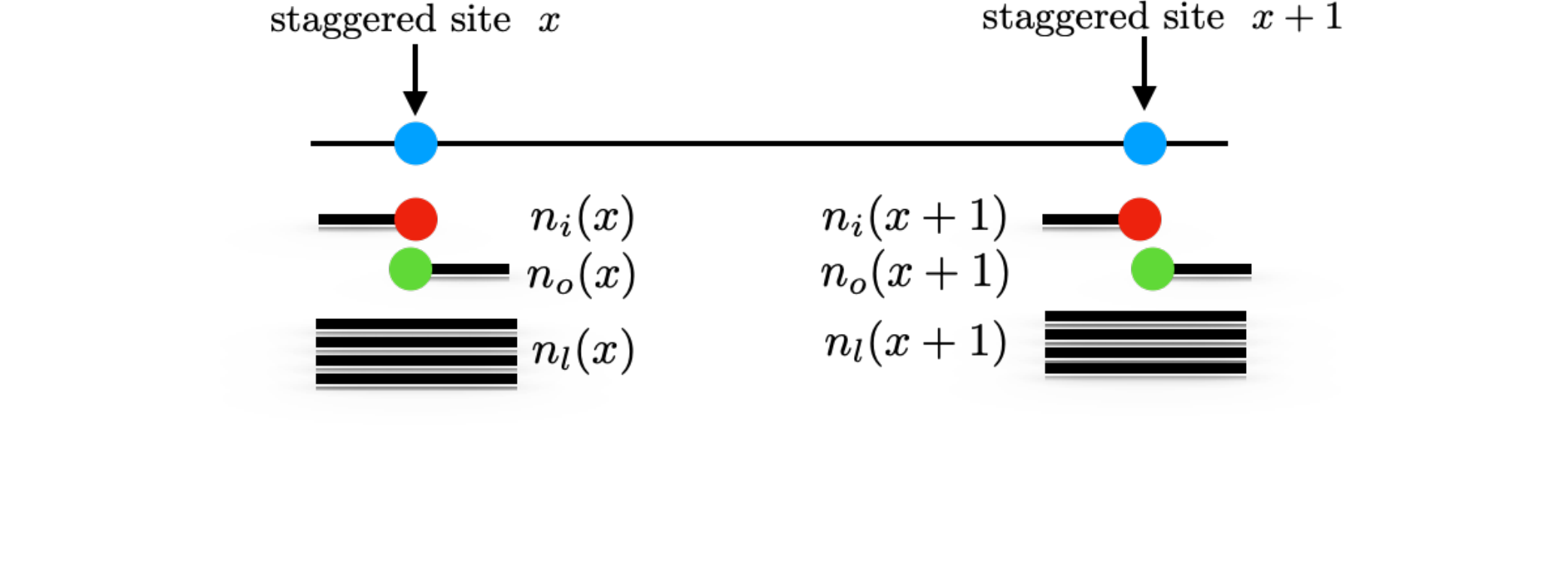}
\caption{Two staggered sites in the LSH formulation on a 1d spatial lattice. Each site  carries three types of operators namely incoming string, outgoing string and flux. The Hilbert space is characterized by the corresponding quantum numbers $n_l,n_i,n_o$ respectively for each and every site of the lattice.}
\label{fig:lsh}
\end{figure}

Within LSH framework,  the gauge invariant and orthonormal LSH basis is characterized by a set of three integers $
n_l(j), n_i(j), n_o(j)$
that satisfies Gauss' law constraint:
\begin{equation}
G^a(j)|n_l(j), n_i(j), n_o(j)\rangle=0,~~~ \forall j , a.
\end{equation}
These three quantum numbers signify loop, incoming string and outgoing string at each site. The allowed values of these integers are given by 
\bea
&&  0\le n_l(j)\le \infty \\
&& 0\le n_i(j)\le 1\\
&&0\le n_O(j)\le 1
\eea
Pictorially, the LSH quantum numbers are illustrated in FIG. \ref{fig:lsh}.
It is clear from the range of the quantum numbers, that $n_l$ is  bosonic excitation,  whereas $n_i,n_o$ are fermionic in nature. However, it is important to note that, unlike fermionic matter field in the original theory, the fermionic operators building the `local string' Hilbert space are SU(2) invariant bilinears of one bosonic prepotential operator and one fermionic matter field, yielding overall fermionic statistics. Hence, the string states contain the information of both gauge field and matter content.  

At this point, we define a set of LSH operators consisting of both diagonal and ladder operators locally at each site as following:
\bea
\label{nl}
\hat n_l|n_l, n_i, n_o\rangle &=& n_l|n_l, n_i, n_o\rangle \\
\label{ni}
\hat n_i|n_l, n_i, n_o\rangle &=& n_i|n_l, n_i, n_o\rangle \\
\label{no}
\hat n_o|n_l, n_i, n_o\rangle &=& n_o|n_l, n_i, n_o\rangle \\
\label{nlpm}
\hat \lambda^{\pm}|n_l, n_i, n_o\rangle &=& |n_l\pm 1, n_i, n_o\rangle \\
\label{nip}
\hat \chi_i^{+}|n_l, n_i, n_o\rangle &=& (1-\delta_{n_i,1})|n_l, n_i+ 1, n_o\rangle \\
\label{nim}
\hat \chi_i^{-}|n_l, n_i, n_o\rangle &=& (1-\delta_{n_i,0})|n_l, n_i- 1, n_o\rangle \\
\label{nop}
\hat \chi_o^{+}|n_l, n_i, n_o\rangle &=&(1-\delta_{n_o,1}) |n_l, n_i, n_o+ 1\rangle\\
\label{nom}
\hat \chi_o^{-}|n_l, n_i, n_o\rangle &=&(1-\delta_{n_o,0}) |n_l, n_i, n_o- 1\rangle 
\eea
In the above set of equations, we have not mentioned explicit site index as these are considered to be defined at a particular site.

One major benefit of using LSH formalism is that, one  no longer needs to solve/satisfy SU(2) Gauss' law (\ref{eq:Ga}) at each site as the basis states are  SU(2) gauge invariant by construction. Note that, for non-Abelian gauge theories imposing Gauss' law is a non-trivial task and that gives rise to a whole range of complications as discussed in \cite{new}. However, the LSH formalism still carries the constraint (\ref{AGL}) that is necessary  to glue SU(2) invariant states residing at neighboring sites to yield original non-local gauge invariant Hilbert space of the theory. In terms of LSH operators, this constraint  (\ref{AGL}) reads as:
 \bea
 \label{AGL_LSH}
 &&\hat n_l(j)+\hat n_o(j)(1-\hat n_i(j))\nonumber \\
 &=& \hat n_l(j+1)+\hat n_i(j+1)(1-\hat n_o(j+1))
 \eea
Comparing each side of (\ref{AGL_LSH}) to that of (\ref{AGL}) upon acting on LSH basis states, we get:
\bea
N_L(j)&=& n_l(j)+ n_o(j)(1- n_i(j)) \label{NL}\\
N_R(j)&=& n_l(j+1)+n_i(j+1)(1-n_o(j+1))\label{NR}
\eea
where, $N_L(j)$ and $N_R(j)$ count bosonic occupation numbers at each end of the link connecting site $j$ and $j+1$.  As mentioned earlier, the bosonic occupation number at each end of a link has contribution coming from fermionic excitation $n_i$ and $n_o$ as well. Pictorially, left and right side of (\ref{AGL_LSH}) and/or (\ref{AGL}) is represented by the number of thick solid lines at left and right end of a link connecting sites $j$ and $ j+1$  in FIG. \ref{fig:lsh}. As in \cite{Raychowdhury:2019iki,new}, definition of a hadronic state in LSH basis is given by  $|n_l=0,n_i=1,n_o=1\rangle$ at one particular site. 

Hamiltonian of the theory, exactly equivalent to the original Hamiltonian (\ref{eq:HKS}) in terms of LSH operators is given by:
\bea
H^{(\rm LSH)}=H^{(\rm LSH)}_E+H^{(\rm LSH)}_M+H^{(\rm LSH)}_I
\eea
where, $H^{(\rm LSH)}_E$ is the electric energy term, $H^{(\rm LSH)}_M$ is the mass term and $H^{(\rm LSH)}_I$ is the matter-gauge interaction term of the Hamiltonian. Explicitly, in terms of LSH operators defined in (\ref{nl}-\ref{nom}), each part of the Hamiltonian is as below:
\bea
\label{HELSH}
H^{(\rm LSH)}_E&=&\frac{g^2a}{2}\sum_n\Bigg[ \frac{\hat n_l(j)+\hat n_o(j)(1-\hat n_i(j))}{2},\nonumber \\
&& \times \left( \frac{\hat n_l(j)+\hat n_o(j)(1-\hat n_i(j)}{2})+1 \right) \Bigg]\\
\label{HMLSH}
H^{(\rm LSH)}_M &=& m\sum_n (-1)^j(\hat n_i(j)+\hat n_o(j)), \\
\label{HILSH}
H^{(\rm LSH)}_I &=&\frac{1}{2a}\sum_n \frac{1}{\sqrt{\hat n_l(j)+\hat n_o(j)(1-\hat n_i(j))+1}}\times \\ \nonumber &&  \Big[ S_o^{++}(j)S_i^{+-}(j+1)
+ S_o^{--}(j)S_i^{-+}(j+1)\\ \nonumber &&
+S_o^{+-}(j)S_i^{--}(j+1)
+S_o^{-+}(j)S_i^{++}(j+1)\Big]\\ \nonumber & \times& 
\frac{1}{\sqrt{ \hat n_l(j+1)+\hat n_i(j+1)(1-\hat n_o(j+1))+1}}.
\eea
Here (\ref{HILSH}) contains LSH ladder operators in the following combinations (suppressing the explicit site index), 
\bea
S_o^{++}&=& \hat \chi_o^+ (\lambda^+)^{\hat n_i}\sqrt{\hat n_l+2-\hat n_i} \\
S_o^{--}&=& \hat \chi_o^- (\lambda^-)^{\hat n_i}\sqrt{\hat n_l+2(1-\hat n_i)} \\
S_o^{+-}&=& \hat \chi_i^+ (\lambda^-)^{1-\hat n_o}\sqrt{\hat n_l+2\hat n_o} \\
S_o^{-+}&=& \hat \chi_i^- (\lambda^+)^{1-\hat n_o}\sqrt{\hat n_l+1+\hat n_o)} 
\eea
and
\bea
S_i^{+-}&=& \hat \chi_o^- (\lambda^+)^{1-\hat n_i}\sqrt{\hat n_l+1+\hat n_i)}  \\
S_i^{-+}&=& \hat \chi_o^+ (\lambda^-)^{1-\hat n_i}\sqrt{\hat n_l+2\hat n_i} \\
S_i^{--}&=& \hat \chi_i^- (\lambda^-)^{\hat n_o}\sqrt{\hat n_l+2(1-\hat n_o)} \\
S_i^{++}&=& \hat \chi_i^+ (\lambda^+)^{\hat n_o}\sqrt{\hat n_l+2-\hat n_o}. 
\eea
The strong coupling ($ga\gg1,ma=$fixed) vacuum of the LSH Hamiltonian is given by:
\bea
n_l(j)&=& 0~~\forall x \nonumber \\
n_i(j)&=&1~,~ n_o(j)~=~1 ~~\mbox{for $j$ odd}\label{SCV_LSH} \\
n_i(j)&=&0~,~ n_o(j)~=~0 ~~\mbox{for ~$j$ even}\nonumber
\eea
It is easy to check that (\ref{SCV_LSH}) satisfies Abelian Gauss law (\ref{AGL_LSH}). 
One should also consider a suitable boundary condition for one dimensional spatial lattice as discussed in detail in \cite{new} as:
\bea
&&\mbox{Open Boundary Condition (OBC):~~}\nonumber \\
&&N_R(0)= l^{\mbox{OBC}}_i \nonumber \\
&&\mbox{Periodic Boundary Condition (PBC):~~}\nonumber \\
&& N_R(0)= N_L(N-1)\equiv l^{\mbox{PBC}}_i.  \nonumber
\eea
where, $N_L,N_R$ are defined in (\ref{NL}) and (\ref{NR}) for the first $(0)$ and last $(N-1)$ site of a $N$ site lattice. $l_i$ can be any positive semi definite integer. Now, one can easily check that, for any gauge invariant state $\prod_{j=0}^{N-1}|n_l(j),n_i(j),n_o(j)\rangle$, the bosonic quantum numbers $n_l(j)$ for all values of $j$ are completely determined by the boundary flux $l_i$ and constraint (\ref{AGL_LSH}) imposed on each and every link of the lattice starting from one end as:
\bea\label{fixnl}
n_l(j)&=&l_i+ \sum_{y=0}^{j-1}\left(n_o(y)-n_i(y)\right)\nonumber \\
 &&-n_i(j)\left(1-n_o(j)\right).
\eea
For OBC, 
any physical state in LSH formalism is completely determined by $(n_i,n_o)$ quantum numbers at each side.  For PBC, the gauge invariant or LSH Hilbert space is characterized by many copies of the same fermionic $(n_i,n_o)$ configurations with different winding number of closed loops, that plays the exact role as the $l_i$ and fixes the $n_l$'s throughout the lattice. We exploit this particular feature in the analog quantum simulation proposal outlined in the present work \footnote{The numerical analysis performed in this work is for OBC in gauge theory as simulating the same in an experiment is easier than that for PBC.}.
Note that, $n_l$ being determined does not mean that we describe a static gauge field theory;  rather, truly relevant or physical gauge degrees of freedom are contained into the $(n_i,n_o)$ excitation of any physical state.



\subsection{Weak coupling approximation}
\label{sec:limit}
\noindent
The strong coupling vacuum of the theory is defined by zero gauge flux i.e $n_l=0$ at all lattice sites. However as one approaches the weak coupling regime, the states containing large amount of bosonic flux do contribute to the low energy spectrum of the theory. In \cite{Raychowdhury:2018osk}, a weak coupling vacuum ansatz was proposed  and justified for the $2+1$ dimensional pure SU(2) gauge theory within prepotential framework. In that proposal, each lattice site contains a large but mean value for the local loop quantum numbers. The $(1+1)$-d version of that ansatz within LSH framework (i.e prepotential + staggered matter) would be equivalent to each site containing  more and more gauge fluxes, i.e $n_l \gg 0$, for all sites  as one approaches the weak coupling limit $g\rightarrow 0$. As discussed before, the incoming flux or boundary flux $l_i$ fixes the bosonic loop quantum numbers $n_l$'s at each of the lattice site for any configurations of $n_i,n_o$ throughout the lattice as per (\ref{fixnl}). Hence, choosing for $l_i \gg 0$ for any finite lattice would result,
\bea
\label{nl-li}
n_l(j)=l_i\equiv n_l ~~~\forall j.
\eea
 We derive the following approximate Hamiltonian that acting on the LSH states on the 1d spatial lattice with the boundary flux $l_i \gg 0$ would result the exact dynamics of the full gauge theory described by the Hamiltonian given in (\ref{HELSH},\ref{HMLSH},\ref{HILSH}).

\bea
\label{wcHELSH}
H^{(\rm approx)}_E
&= & \frac{g^2a}{2} \sum_j\Bigg[\frac{\hat n_l(j)}{2} \frac{\hat n_l(j)}{2} \Bigg]\\
\label{wcHMLSH}
H^{(\rm approx)}_M &=& m\sum_j (-1)^j(\hat n_i(j)+\hat n_o(j)) \\
\label{wcHILSH}
H^{(\rm approx)}_I &= &\frac{1}{2a}\sum_j  \Big[ \hat{\chi}_o^{+}(j)\hat{\chi}_o^{-}(j+1)
+ \hat{\chi}_o^{-}(j)\hat{\chi}_o^{+}(j+1) \nonumber \\ &&
+\hat{\chi}_i^{+}(j)\hat{\chi}_i^{-}(j+1)
+\hat{\chi}_i^{-}(j)\hat{\chi}_i^{+}(j+1)\Big]
\eea
The derivation of the approximated Hamiltonian given above is detailed in Appendix \ref{App:wcHamiltonian}.
For $n_l\simeq l_i \gg 0$ at $g\to 0,a\to 0$, the weak coupling LSH Hamiltonian gives an accurate description of low energy spectrum of the continuum. In this limit, the Abelian Gauss law constraint (\ref{AGL}) is automatically satisfied as (\ref{NL}) and (\ref{NR}) effectively become equal.

We present the details of an atomic quantum simulation scheme to simulate this approximated weak coupling Hamiltonian in next section.

The systematic correction to recover the full Hamiltonian is also discussed in Appendix \ref{App:wcHamiltonian} that is to be taken into account in order to improve upon this particular quantum simulation proposal. Systematic generalization of this proposal in a series of future works would finally lead to a complete and scalable quantum simulator for non-Abelian gauge theories.   


\section{Atomic Hamiltonian :  Hubbard Model on a Bipartite Lattice }
\label{sec:atomic}
\begin{figure*}[t]
\includegraphics[width=0.70\textwidth]{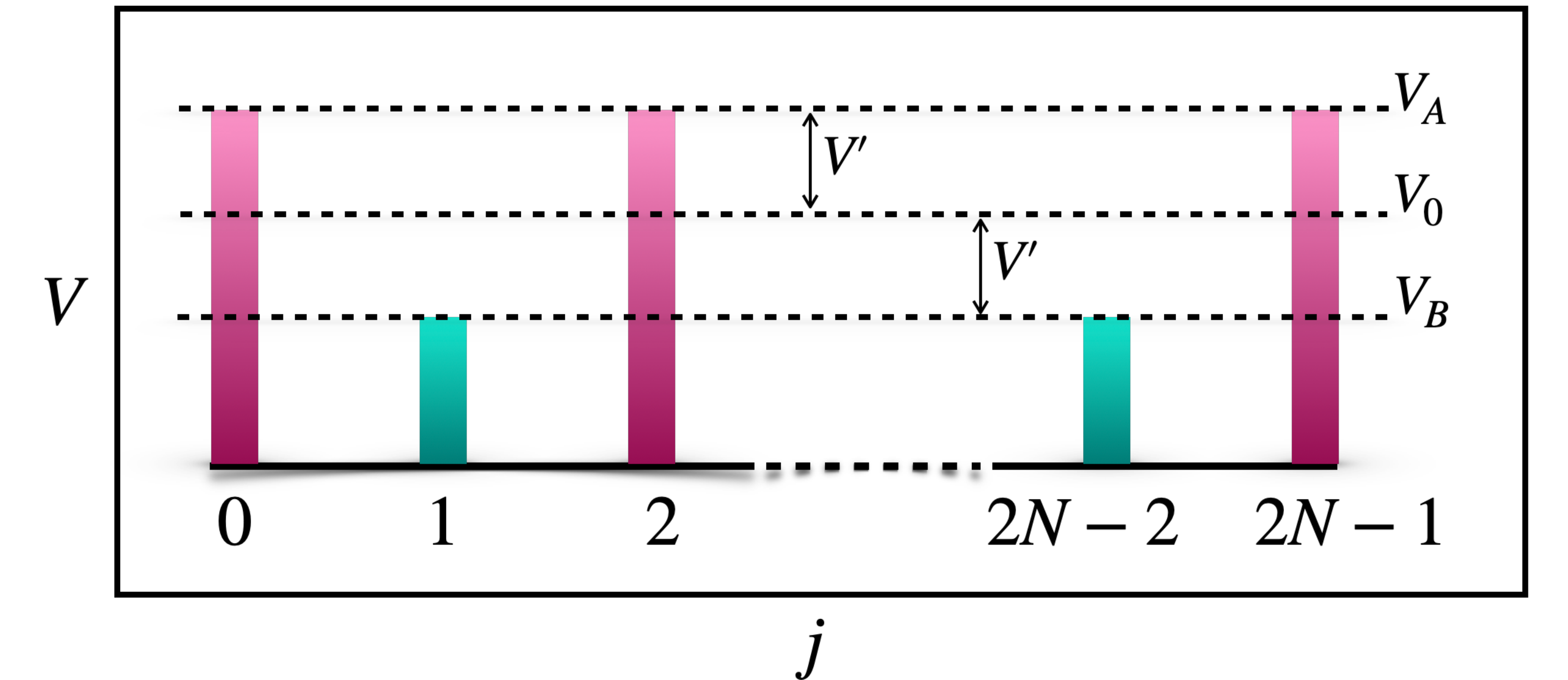}
\caption{Structure of the two-color lattice}
\label{fig:KP}
\end{figure*}

\noindent
We consider a Fermi-Hubbard model in a one-dimensional lattice. The lattice is a superposition of a primary lattice and a secondary lattice. The primary lattice has the form 
\bea
V_1(x)= -\sum_j V_L\delta(x-x_j),
\eea
and can also act as the trapping potential. Here $V_L$ marks the primary lattice depth. In the secondary lattice,  the lattice depth is $V_A$ and $V_B$ on alternating sites. The secondary lattice thus is a Kronig-Penney type two-color lattice: the potential consisting of  $\delta$-combs  of two different strengths: 
\bea
V_2(x)= \sum_{j=\mbox{\scriptsize{odd}}} V_A\delta(x-x_j)+\sum_{j=\mbox{\scriptsize{even}}} V_B\delta(x-x_j)
\eea 
The structure of the secondary lattice is shown in Fig. \ref{fig:KP}. We introduce two new parameters : $V_0=(V_A+V_B)/2$ and $V'=(V_A-V_B)/2$.

The fermionic atoms can belong to either of its two accessible hyperfine states : we denote them by the symbols $|\uparrow  \rangle $ and $|\downarrow  \rangle $ respectively;  $\psi_\uparrow(x)$ and $\psi_\downarrow(x)$ being the corresponding field operators.
The Hamiltonian can be written as 
\bea
H=H_0+H_{\mbox{\scriptsize{int}}}.
\eea
Here $H_0$ is the non-interacting part of the Hamiltonian, and $H_{\mbox{\scriptsize{int}}}$ is the fermion-fermion interaction. 
\bea
H_0=\int  \big( \psi_\uparrow^\dagger(x) \mathcal{H}_0(x)\psi_\uparrow(x)+ \nonumber \\
\psi_\downarrow^\dagger (x)  \mathcal{H}_0(x)\psi_\downarrow(x)\big ) dx,
\eea 
where 
\bea
\mathcal{H}_0(x)=\mathcal{H}_1(x)+\mathcal{H}_2(x)
\eea 
\bea
\mathcal{H}_1(x)=\dfrac{-\hbar^2}{2m}\dfrac{\partial^2}{\partial x^2 }+V_1(x)
\eea 
is the contribution from the kinetic energy and the primary lattice potential, same for all the sites. The contribution from the secondary lattice is 
\bea
\mathcal{H}_2(x)=V_2(x)=V_0+V'
\eea 
for  odd sites. And 
\bea
\mathcal{H}_2(x)=V_2(x)=V_0-V'
\eea 
for even sites. 

In the low-energy scattering regime, the atoms usually interact via s-wave scattering. The corresponding coupling constant is given by $$g_0=\dfrac{4\pi \hbar^2 a_s}{m}$$ $a_s$ being the scattering length. 
The interacting part of the Hamiltonian is given by: 
\bea
H_{\mbox{\scriptsize{int}}}=g_0 \int  \psi_\uparrow ^\dagger(x) \psi_\downarrow^\dagger (x) \psi_\downarrow(x)\psi_\uparrow(x) dx.
\eea 
Now, if the  lattice potentials are sufficiently deep, the field operators can be expanded in terms of single-particle Wannier functions, localized to each lattice site:
\bea
\psi_\sigma(x)=\sum_j c_{\sigma}(j) \mathcal{W}(x-x_j); \hspace{20 pt} \scriptsize{\sigma=\uparrow,\downarrow},
\eea 
where $c_{\sigma}(j)$ is the fermionic annihilation operator for spin index $\sigma$ and site $j$. The single-particle Hamiltonian can be written as:  
\bea
H_0=H_\epsilon+H_{V_0}+H_{V'}+H_{\mbox{\scriptsize{hopping}}}.
\eea
Here, each term can be expressed using the fermionic creation and annihilation operators $c_{\sigma}^\dagger (j)$ and $c_{\sigma}(j)$; and also the number operators $c_{\sigma}^\dagger (j)c_{\sigma} (j)=\mathcal N_{j\sigma}$. 

The first term is : 
\bea H_\epsilon = \sum_j \epsilon_j \mathcal N(j)
\eea 
where the parameter $\epsilon_j$ is given by 
\bea
\epsilon_j= \int \mathcal{W}(x-x_j)\mathcal{H}_1(x)\mathcal{W}(x-x_j) dx
\eea 
and  $\mathcal N(j)=\mathcal N_{\uparrow}(j)+\mathcal N_{\downarrow}(j)$ is the total number of fermions in site $j$. In the subsequent part of the paper, we neglect the $H_\epsilon$ term because if simply gives a constant energy shift.

As for the terms arising from the secondary lattice : 
\bea 
H_{V_0}=\sum_j V_0 \mathcal N(j)
\eea 
and 
\bea 
H_{V'}=V'\sum_{j=\mbox{\scriptsize{odd}}} \mathcal N(j)- V'\sum_{j=\mbox{\scriptsize{even}}} \mathcal N(j) 
\eea 
$H_{V_0}$ is a constant, too, but we keep this, in order to make a direct correspondence with the reduced LSH Hamiltonian. 
 
The hopping term, which represents the tunneling  between sites is given by : 
\bea
H_{\mbox{\scriptsize{hopping}}}=- \sum_j t_{i,j}\big(  c^\dagger_{\uparrow}(j) c_{\uparrow}(i)
+c^\dagger_{\downarrow}(j) c_{\downarrow}(i)\big).
\eea 
Tunneling to next-nearest neighbors is usually suppressed by one order of magnitude, in comparison  with the nearest neighbor tunnelling. So we consider hopping between adjacent sites only. 
The tunneling rate from site $j$ to $(j+1)$ is given by the matrix element 
\bea
t_{j,(j+1)}=- \int \mathcal{W}(x-x_j)\mathcal{H}_{0}\mathcal{W}(x-x_{j+1}) dx.
\eea
As for the interaction between up-spin and down-spin fermions sharing the same site
\bea
\label{HU}
H_{\mbox{\scriptsize{int}}}=u \sum_j \mathcal N_{\uparrow}(j)   \mathcal N_{\downarrow}(j).
\eea 
Here the on-site interaction matrix element  is given by 
\bea
u=g_0\int |\mathcal{W}(x-x_j)|^4 d x.
\eea
The total Hamiltonian thus translates to
\bea 
\label{HamFH}
H=&&-\sum_j t_{j,(j+1)} \big( c^\dagger_{\uparrow}(j) c_{\uparrow}(j+1)
+c^\dagger_{\downarrow}(j) c_{\downarrow}(j+1)\big) \nonumber \\ && +u \sum_j \mathcal N_{\uparrow}(j)   \mathcal N_{\downarrow}(j)  +\sum_j V_0 \mathcal N(j)\nonumber \\
 &&+{V'}\sum_{j=\mbox{\scriptsize{odd}}}\mathcal N(j)- V'\sum_{j=\mbox{\scriptsize{even}}}\mathcal N(j).
\eea

If the hopping $-t$ is a constant throughout the lattice,  this model essentially is  a 1D Hubbard model  with alternating potential, often termed as the `` ionic Hubbard model", defined on a bipartite lattice. Here, in addition to a site-independent hopping $-t$ and the on-site interaction  $u$, there is a difference in the energy offset $2V'$ between sublattice \textbf{A} and sublattice \textbf{B}. This model was originally proposed to study transitions in organic crystals \cite{nagaosa}, and later, found application in the studies of  ferroelectric transitions \cite{egami}.  In the recent past, this model  has been experimentally realized \cite{messer} in a system of ultracold atoms. So we consider this to be a very suitable candidate to simulate lattice gauge theories. 

At half-filling, the ionic Hubbard model is capable of describing a  band-insulator \cite{fabrizio1999band}. However, this  model has a rich phase diagram, and at higher inter-atomic interaction strengths, can support transitions to different states, including  Mott insulator \cite{fabrizio1999band}, correlated insulator \cite{kampf2003nature, bag2015phase}, AFM insulator and half-metal  \cite{bag2015phase} phases; and certain combinations of $u$ and $V'$ can even lead to superfluidity \cite{samanta2016superconductivity}. As we will see in the later part of this work,  we will have to carefully choose our parameters such that the entire dynamics remains confined to a single paramagnetic phase in order to mimic the dynamics of gauge theory. 
  
\section{Simulating and Simulated Hamiltonian and their parameters}
\label{sec:parameter}
\noindent
We are now in a position to compare the weak coupling LSH Hamiltonian and the atomic Hamiltonian. For a particular site $j$,  we make the following identification :
\bea
n_i(j)=\mathcal N_{\uparrow}(j); \hspace{20 pt} n_o(j)=\mathcal N_{\downarrow}(j)\\
\chi_i^{+}(j) =c^\dagger_{\uparrow}(j); \hspace{20 pt} \chi_i^{-}(j) =c_{\uparrow}(j)\\
\chi_o^{+}(j) =c^\dagger_{\downarrow}(j); \hspace{20 pt} \chi_o^{-}(j) =c_{\downarrow}(j)\\
m=V'
\eea 
Also, the magnitude of $V_0$ has to be chosen to be mapped to electric part of the gauge theory Hamiltonian for a particular $n_l$, fixed by the open boundary condition. 

The electric term of approximated LSH Hamiltonian is mapped to: 
\bea 
\label{EtoV0}
H_E^{(\mathrm{approx})} \rightarrow \sum_j V_0 \mathcal N(j). 
\eea 
Similarly, the potential $V'$ is fixed by the mapping:
\bea 
\label{MtoV'}
H_M^{(\mathrm{approx})} \rightarrow {V'}\sum_{j=\mbox{\scriptsize{odd}}}\mathcal N(j)- V'\sum_{j=\mbox{\scriptsize{even}}}\mathcal N(j) 
\eea 
and the hopping terms are identically related as,
\bea 
\label{Itot}
H_I^{(\mathrm{approx})} &\rightarrow& -t \sum_j \big(  c^\dagger_{\uparrow}(j) c_{\uparrow}(j+1)
+c^\dagger_{\downarrow}(j) c_{\downarrow}(j+1)\big) \nonumber \\ && + \mbox{h.c}. 
\eea 
Note that, there is no term in the weak coupling LSH Hamiltonian that corresponds to the on site interaction term (\ref{HU}). So, in the limit $u\rightarrow 0$, one would have a complete mapping between atomic system and weak coupling limit of gauge theory.  

We would like to point out that although the LSH Hamiltonian contains explicit bosonic modes $n_l(j)$, these are actually non-dynamical in the weak coupling approximation as discussed before, and hence we do not keep actual bosons in the atomic system. Instead, we incorporate the effect of these bosons in the potential itself, in the form of a constant energy shift. This enables us to i) keep $n_l$ uniform for each site, and ii) ensure that the bosonic and the fermionic modes are completely decoupled: as there remains no chance of any boson-fermion scattering. 


Moving away from weak coupling approximation, the electric part becomes dominant and hence correction to the same becomes most important. The complete correction to the electric part of Hamiltonian is given in (\ref{HEmf}) within mean field ansatz.  The on-site interaction in the atomic Hamiltonian, which does not have an equivalent in the approximate LSH Hamiltonian, can be tuned to recover the exact  contribution of (\ref{HELSH}) within the mean field ansatz of LSH Hamiltonian given in (\ref{HEmf}). 

As discussed in Appendix \ref{App:wcHamiltonian}, the correction term to be added to the weak coupling approximated electric Hamiltonian (\ref{wcHELSH}) to yield full electric Hamiltonian (\ref{HELSH}) in the bulk limit is given by:
\bea
\label{corr_HE}
\Delta H_E^{(\rm LSH)} = \frac{g^2 a }{2}\frac{N}{4} \left(\frac{n_l}{2}+\frac{3}{4}\right)
\eea 
Now, for a Hubbard model at half-filling, all the four accessible states  $| 0\rangle$,  $|\uparrow\rangle $, $| \downarrow \rangle $ and $|\uparrow \downarrow\rangle $ are equally likely as long as the system remains in the paramagnetic phase. So, $N_{|\downarrow\rangle }$ (the number of sites belonging to state  $|\downarrow\rangle$) $\approx N/4$, $N$ being the total number of lattice sites. Similarly $N_{|\uparrow \downarrow \rangle }$, the number of sites with doublons, would be $\approx N/4$, too  and those many configurations contribute to  (\ref{HU}). 
Hence,  one can utilize the on-site interaction term to recover the exact correction term,
\bea u\sum_j \mathcal N_{\uparrow}(j)   \mathcal N_{\downarrow}(j)=u\frac{N}{4} ~\longrightarrow ~\Delta H_E^{(\rm LSH)} \label{UtoEcorr}\eea 

\begin{figure*}[t]
\includegraphics[width=0.79\textwidth,height=0.8\textwidth]{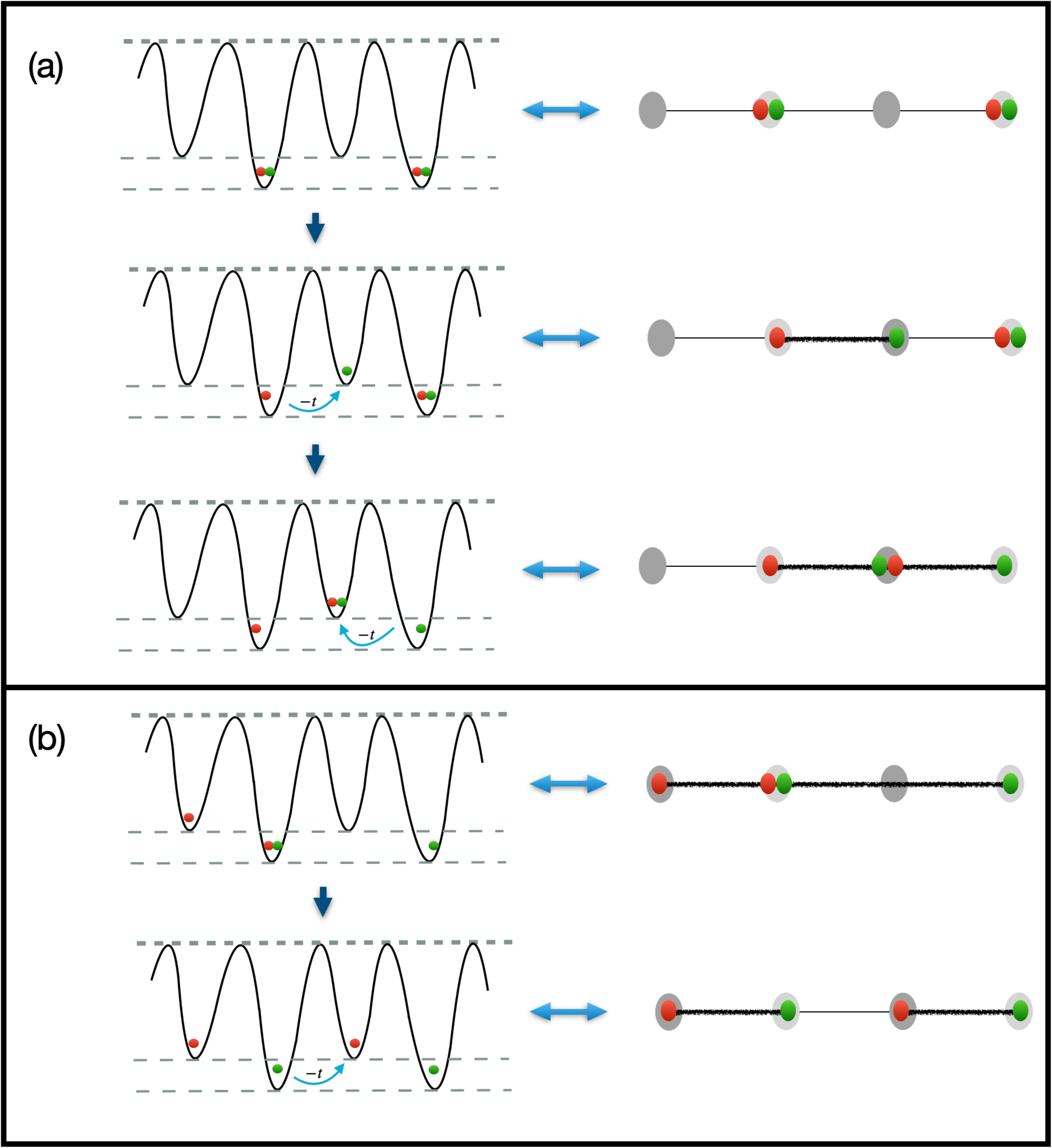}
\caption{Cartoon representation of dynamics of 1d ionic Hubbard model mimicking that of SU(2) lattice gauge theory in one spatial dimension. (a) Initial state: fully filled odd sites and empty even sites mimicking the strong coupling vacuum consisting of no particles ($n_i=0,n_o=0$ on even sites) and no antiparticles ($n_i=1,n_o=1$ on odd sites). Under Hamiltonian evolution, one atom hops from an odd site to neighboring even site in Hubbard model, that mimics creation of a particle antiparticle pair at two neighboring staggered site of the gauge theory, connected by one unit of flux to form a gauge singlet string configuration. One further hopping as shown in the figure mimics the dynamics in gauge theory as elongation of the string and creation a baryon on one site. In these three states, the total number of particle (antiparticle) for the gauge theory are respectively $0,1,2$. (b) Ionic Hubbard model dynamics is mimicking string breaking dynamics of gauge theory. Starting from a string of length 3 unit, pair production occurs and the initial string breaks into two smaller strings.} 
\label{fig:dynamics_cartoon}
\end{figure*}

Note that, (\ref{UtoEcorr}) is exact, only in the bulk limit, i.e $N\gg2$ lattice size and in that limit, we have the maximum overlap of the (mean field approximated) strong coupling lattice gauge theory to  Fermi-Hubbard Hamiltonian as
Then 
\bea
H^{(\mathrm{mLSH})}_E &\longrightarrow& H_{V_0}+H_{\mbox{\scriptsize{int}}}\\
H^{(\mathrm{LSH})}_M &\longrightarrow& H_{V'}\\
H^{(\mathrm{approx})}_I &\longrightarrow& H_{\mbox{\scriptsize{hopping}}}
\eea 
provided we fix $V_0$ and $u$ such that the system, staying in the desired phase, mimics the dynamics of gauge theory as shown in Fig. \ref{fig:dynamics_cartoon}.

The correction to the approximate interaction Hamiltonian is negligible in weak coupling regime, and also insignificant in the strong coupling limit. In this work, we do not consider any correction to the interaction term.

We now, explicitly calculate the parameters of the atomic Hamiltonian to simulate desired gauge theory dynamics, that has to be tuned in the experiment. First, we scale the gauge theory Hamiltonian to be dimensionless in order to make comparison with that of atomic system.

\subsection{Scaling of the gauge theory Hamiltonian}
\label{Sec:scalingHamiltonian}
\noindent

It is convenient to scale the Hamiltonian $H^{\mbox{(KS)}}$ given in (\ref{eq:HKS}) as per \cite{Hamer:1981yq}, so as to make it dimensionless:
\bea 
\label{scaledHwc}
\tilde H&=& \frac{2}{g^2a} H^{(\mbox{KS})} \nonumber \\
&=& \underbrace{\sum_{j}{E^2}(j)}_{\tilde H_E} +\underbrace{\mu_0 \sum_{j}(-1)^j\left[\psi^{\dagger}(j)\cdot \psi(j) \right]}_{\tilde H_M}\nonumber \\
&&+ \underbrace{x_0\sum_{j}\left[\psi^{\dagger}(j) U(j) \psi(j+1)+{\rm h.c.} \right]}_{\tilde H_I}.
\eea
Here, $x_0=\frac{1}{g^2a^2}$ and $\mu_0= 2\sqrt{x_0}\dfrac{m}{g}$ are dimensionless coupling constants of the theory.
Evolving this $\tilde H$ with scaled time (from zero to $\tilde \tau$)
\bea \tilde \tau=\frac{\tau_{\mathrm{gauge}}}{x_0} \label{scaledtime_gauge}\eea is due to the unitary operator:
\bea
\mathcal U(\tilde \tau)&=& \exp\left({-i\tilde H \tilde \tau}\right) \nonumber \\
&=&\exp\left(-i\frac{2}{g^2a}H^{\mathrm{(KS)}} g^2a^2 \tau_{\mathrm{gauge}}\right) \nonumber \\
&=& \exp\left(-i2aH^{\mathrm{(KS)}}\tau_{\mathrm{gauge}}\right).
\label{unitarysc}
\eea
Here, $2aH^{\mathrm{(KS)}}$ is another scaled Hamiltonian with dimensionless parameters given by,
\bea
\label{scaledHsc}
2aH^{(\mathrm KS)} &=& \frac{1}{x_0}\sum_{x}{E^2}(x)\nonumber \\
&&+ 2\frac{m}{g}\frac{1}{\sqrt{x_0}}\sum_{j}(-1)^j\left[\psi^{\dagger}(j)\cdot \psi(j) \right] \nonumber \\
&& +\sum_{j}\left[\psi^{\dagger}(j) U(j) \psi(j+1)+{\rm h.c.} \right].
\eea
The strong coupling limit is defined for $x_0\rightarrow 0$, where the interaction part of Hamiltonian become less dominant as evident from both (\ref{scaledHwc}) and (\ref{scaledHsc}), whereas in the weak coupling limit defined at $x_0\rightarrow \infty$, interaction part of the Hamiltonian becomes the most important term that cannot be treated perturbatively. These scaling rules work equivalently on the LSH Hamiltonian defined in (\ref{HELSH},\ref{HMLSH},\ref{HILSH}) as the LSH Hamiltonian is exactly equivalent to the original Kogut-Susskind Hamiltonian.

\subsection{Weak coupling regime of gauge theory:}
\label{sec:ws-scaling}
\noindent
For the Hamiltonian given in (\ref{scaledHwc}),
we consider the bosonic loop quantum number to take the average value
\bea
n_l \approx \mathcal O(10^p) \Rightarrow \tilde H^{(\mathrm{approx})}_E \approx \mathcal O(10^{2p}).
\eea
For a comparative mass and interaction contribution of the Hamiltonian, i.e.
\bea
 \tilde H^{LSH}_M &\approx &  \mathcal O(10^{2p})
\\
\& \tilde H^{(\mathrm{approx})}_I &\approx & \mathcal O(10^{2(p+p')}) ,
\eea
is obtained for the following scaling of the parameters: \bea  \frac{m}{g} \approx \mathcal O(10^{p-p'})~~\sqrt{x_0} \approx \mathcal O(10^{p+p'}) ~~,\forall p'\in \mathbb Z_+.\eea
The exact values of the dimensionless parameters of gauge theory can be taken as:
\bea
\label{nlp}
n_l&=& \tilde n_l \times 10^p \\
\mu_0&=& \tilde \mu_0 \times 10^{2p}\\
x_0 &=& \tilde x_0 \times 10^{2(p+p')}, ~~\forall \mbox{ integer } p,p' \label{pp'}\\
&=& 10^{2p} \mbox{ for the choice, } p'=0, \tilde x_0=1.
\eea
Now, the dynamics of this scaled Hamiltonian $\tilde H$ in (\ref{scaledHwc}),  is to be simulated by the simulating Fermi-Hubbard Hamiltonian given in (\ref{HamFH}) in the time scale $\tilde \tau$ as defined in (\ref{scaledtime_gauge}), such that
\bea
\label{dynwc}
\exp({-i\tilde H\tilde\tau})\longrightarrow \exp({-iH\tau})
\eea
where, $H$ is the atomic Hamiltonian given in (\ref{HamFH}) with the parameters: 
\bea
\label{wcfix1}
V'&=& \tilde \mu_0  \\
V_0 &=& \frac{1}{4}\left( \tilde n_l^2+2\tilde n_l \right) \\
u &=& 0 \label{wcfix2} \\
t &=& -1 \label{t2x0}.
\eea
Here, all the parameters are fixed in units of `$t$'. The only choice that we have made in setting the parameters is $p'=0$ in (\ref{pp'}). Gauge theory with a nonzero $p'$ can be equivalently simulated by the same atomic system with tuning $V'$ to smaller values $\tilde \mu_0 \times 10^{-2p'} $  in an experiment. This will access all mass values of gauge theory in the quantum simulation protocol. 

\subsection{Strong coupling regime of gauge theory:}

We consider the scaled Hamiltonian in (\ref{scaledHsc}) in strong coupling regime $x_0<1$. 
As discussed earlier, the bulk limit of the Fermi-Hubbard Hamiltonian  in the paramagnetic phase will correspond to the exact mean field electric term (\ref{HEmf}) and mass term  (\ref{HMLSH}). Although the interaction term is approximated,  will not make major difference in spectrum and/or dynamics as $x_0\rightarrow 0$ as it is less dominant compared to diagonal terms. 
Likewise, weak coupling regime, we fix the boundary condition $l_i$ to be a fixed integer, but is of $\mathcal O(1)$. We map the gauge theory Hamiltonian to Fermi-Hubbard Hamiltonian with parameters 
\bea
\label{scfix1}
V'&=& 2\frac{1}{\sqrt{x_0}}\cdot\frac{m}{g} \\
V_0 &=& \frac{1}{{x_0}}\cdot\frac{l_i}{4}\left( l_i+2 \right) \label{fixV0}\\
u &=&  \frac{1}{{x_0}}\cdot\frac{1}{4}\left( 2l_i+3 \right) \label{scfix2}\\
t &=& -1. \label{t2a}
\eea
Here also, all the parameters are fixed in units of `$t$'. It is clear from the above relations, for a fixed value of $l_i$, smaller values of $x_0$ require larger $V_0/t$ and $u/t$ for the atomic system. However, we will have to be careful to remain in the same paramagnetic phase such that our analysis of compensating errors in electric Hamiltonian from the uniform potential are well compensated by the self interaction term. For this purpose, i.e in order to keep $u/t$ below the critical point for paramagnetic-ferromagnetic phase transition one can not really expect to simulate $x_0 \rightarrow 0$ under the present scheme. However, one  can  simulate $x_0<1$ as well as $x_0=1$ besides accurately simulating intermediate coupling range $x_0\approx 10-100$ as will be demonstrated in the numerical analysis.   

Likewise the weak coupling case, the  simulating and simulated dynamics are comparable up to a factor 
\bea \tau_{\mbox{atomic}}=2a\times  \tau_{\mbox{gauge}}. \eea
where, a is small but finite in strong coupling limit. 

In the next section, we propose the precise experimental set-up that is close to already performed experiments for Ionic-Hubbard model following the above mentioned scheme, where strong coupling regime of lattice gauge theory dynamics is mapped to ionic Hubbard model with $u/t>1$, whereas the weak coupling regime is mapped to the same with $u/t\approx 0$. 

\section{Experimental Realization}
\label{sec:exp}       
\noindent

The experimental scheme calls for the realization of 1-dimensional  Fermi-Hubbard  model  in  a  bipartite   lattice. In the recent past, the ionic Fermi-Hubbard model was experimentally realized in a honeycomb lattice \cite{messer}, and its bosonic counterpart was implemented on a bipartite chequerboard lattice \cite{liberto}. Also, a 1-dimensional Fermi-Hubbard model was implemented in a experiment by Scherg et al.\cite{scherg}.  Both \cite{messer} and \cite{scherg} used a degenerate  gas  of  fermionic $^{40}\rm{K}$ of numbers $\approx 10^5$ and  $10^4$ respectively.  We propose that a combination of these two methods can successfully yield a 1-dimensional Hubbard model with alternating lattice potentials.

\subsection{Proposed set-up}
The interference pattern of two counter propagating lasers is used to create an optical lattice. The lattice depth is proportional to the intensity of the laser beam and is measured in units of the recoil energy $E_R$. 

In the experiment by Messer et al.\cite{messer}, first a regular honeycomb lattice was created, and that fixed the hopping parameter $t$ on each bond. Next a staggered energy offset of $\Delta$ was independently applied between sites of $\textbf{A}$ and $\textbf{B}$ sublattices. In our 1-dimensional structure, an equivalent would be to set up the primary lattice with lattice depth $V_L$ :  $$V_1(x)=-V_L \rm{cos}^2(x)$$ and  superpose that with $$V_2(x)=V_0$$ on each site. This fixes the hopping parameter $t$. Then, on top of it,  energy offsets  $V'$ and $-V'$ can be independently applied on the odd sites and even sites respectively, so that $V_2(x)=V_0+V'$ for odd sites, and $V_2(x)=V_0-V'$ for even sites.

Just like the hopping $t$, the on-site interaction $u$, too depends on the lattice depth. However, $u$ can be independently controlled as well, by means of Feshbach resonance. As for the two fermionic states, any two hyperfine states of a particular atom can be employed. In  \cite{scherg},  the hyperfine states
\bea|\uparrow\rangle &=&|F=-9/2; m_{\rm{F}}= -9/2\rangle  \nonumber \\  ~~|\downarrow\rangle &=&|F=-9/2; m_{\rm{F}}= -7/2\rangle\label{comb1} \nonumber \eea of ultracold $^{40}\rm K $ atoms were used. In \cite{messer}, in addition to the above, the combination   \bea|\uparrow\rangle &=&|F=-9/2; m_{\rm{F}}= -9/2\rangle\nonumber \\  ~~|\downarrow\rangle &=& |F=-9/2; m_{\rm{F}}= -5/2\rangle \nonumber \eea was also employed in order to obtain desired range of $u$. 

In \cite{messer}, the ionic Hubbard model was studied on a honeycomb lattice. In contrast, our model requires the implementation of the ionic Hubbard model in a simple one-dimensional geometry. Regarding the dimensionality of the system, it may be recalled that in the recent past, ultracold atom experiments have successfully confined bosonic and fermionic atoms  to one dimension (1D). The basic idea is to tightly confine the particles in two transverse directions, and make them weakly confined in the axial direction. Thus, their motion in the transverse directions are completely frozen. So effectively, these are quasi-1D systems. 

For example, in our proposed set up, suppose both $V_y$ and $V_z$, the potentials in the transverse directions, are kept fixed at a large value (Like, $33 E_R$ as in  \cite{ronz}, or $42E_R$ as in \cite{spon}). $V(x)$, The lattice depth in the axial direction is governed by both $V_L$ and $V_0$, and the final depth is kept in a range of $5 E_R -12 E_R$. We note that in Hubbard model experiments, the potentials are to be deep enough ($V\geq 5E_R$) so that the single-band description of Hubbard model remains valid. On the other hand, $V(x)$ cannot be as deep as the potentials in the transverse direction, so as to restrict the dynamics in 1-dimension only. The hopping parameter $t$ is a function of the lattice depth, and can be estimated using the Wannier functions \cite{bloch}.  

The actual lattice depth is given by $|V_L-V_0|$, so different combinations of $V_L$ and $V_0$ can result in the same lattice depth. This offers a tremendous advantage in the experimental pursuit, as the same optical lattice can be assumed to be split in different pairs of $V_L$ and $V_0$ : allowing one to explore a wide range of $V_0$ values (that, in turn, enables one to access a wide range of  $x_0$ and/or $l_i$ as per (\ref{fixV0})).  It is to be noted that both $V_L$ and $V_0$ are theoretical parameters in the model that leads to constant shifts in the energy only : bearing no effect on the dynamics of the fermions.

Accordingly, we consider two configurations : \begin{enumerate}\item[(i)]  $V_L= 6 E_R$ and $V_0= 0.5 E_R$ and \item[(ii)] $V_L= 6.5 E_R$ and $V_0= 1E_R$.\end{enumerate} In both the cases,  the resultant uniform lattice depth is $5.5 E_R$ for all the sites. This results in a hopping $t=0.057 E_R$. 
The combinations we have mentioned translate to 
\bea
\mbox{(i)}~~ V_0\approx 8.75 t~~\mbox{ and~~~ (ii)}  V_0\approx 17.5 t \nonumber \eea
respectively. In addition, an offset of $V'$ and $-V'$ is independently applied on the odd and even sites. In our scheme, we choose $V'=1.6 t$ and stick to this value in all our numerical simulations. The on-site interaction $u$ can be controlled by applying a Feshbach field. 

To simulate the weak coupling limit, we restrict ourselves to the weakly interacting atomic limit : $u/t << 1$, and  choose $u=0.1 t$. On the other hand, simulation of the strong coupling limit calls for the realization of  the strongly interacting atomic limit : $u/t \gg 1$, and we choose $u\approx 5.5 t$. We note that  these $V'/t$ and $u/t$ values comfortably fall in the parameter regimes accessed in recent experiments \cite{scherg, messer, schreiber}. 

\subsection{Initial state preparation: }
The initial state has to be prepared in a Charge-density-wave (CDW) configuration where all the odd sites are occupied by the fermionic particles and the even sites are completely empty. This can be done using some sort of filtering sequence in the experiment.  For example, in \cite{schreiber, aidel}, this was achieved by superposing the primary lattice (with wavelength $\lambda$) with an additional long lattice (with wavelength $2\lambda$) in the following way:  
\bea 
V(x) =-V_l(\cos^2(k x/2 +\phi))-V_s \cos^2(k x ) 
\eea
with $k=2\pi/\lambda$.

The lattice depths $V_l$ and $V_s$ and the relative phase $\phi$ can be adjusted independently. Here $V_s$ stands for the depth  of the original (and short) lattice with $V_s=|V_L-V_0|$; and $V_l$ is the depth of the additional long lattice.  This long lattice is  utilized  during the preparation of the initial CDW state. Initially, the long lattice is made quite deep (like, $20E_R$, as in \cite{schreiber}), and the short lattice  is  ramped up to that depth at a non-zero relative phase $\phi$ to create a tilted lattice of double wells. Now it is so arranged that the odd sites host lower energy wells than the even sites, and it is possible to load all atoms in the odd sites only. The tilt offsets are made sufficiently large so that the particles cannot escape from the odd sites and tunnel to the even sites. After loading all the atoms, the longer lattice is switched off, and the short lattice is ramped down to its desired final value (In our case, $5.5 E_R$). The offset $V'$ and $-V'$ is added to the odd sites and even sites respectively, to create the bipartite structure.  Now tunneling is possible between adjacent sites, and the the dynamics begins. 
\subsection{Observing the dynamics}
The observable can be defined as the population imbalance $P$ between the even sites and odd sites, defined as 
\bea
\label{defP}
P=\dfrac{N_e-N_o}{N_e+N_o}.
\eea
Here $N_e$ is the total number of atoms in the even sites, and $N_o$ is the total number of atoms in the odd sites. 

 The time evolution of the parameter $P$ is to be studied in order to visualize the particle number dynamics of gauge theory.  A site-resolved technique is thus needed to determine the number of atoms on even and odd lattice sites separately. In \cite{schreiber, aidel}, a band-mapping scheme was successfully employed using the long lattice. Once the desired time evolution in the primary (short) lattice is over, the long lattice is introduced again to create the tilted lattice, and tunneling stops. The phase $\phi$ is chosen such a way that the odd sites constitute the lower wells in the array of double wells. Now the population distribution across the odd and even sites gets sealed. Next, the depth of the long lattice is ramped to a much higher value : and the atoms in the even sites get transferred to the third Bloch band of the superlattice. Atoms in the odd sites remain in the first band . The density profile in the different bands can be obtained using Time-of-flight (TOF) images and absorption imaging \cite{schreiber}. 



\section{Simulated dynamics and observables}
\label{sec:dynamics}
\noindent
We present numerical analysis of our proposal and demonstrate the comparison between the simulating and simulated spectrum as well as dynamics.
\subsection{Spectrum comparison:}

\begin{figure*}[t]
\includegraphics[width=0.99\textwidth]{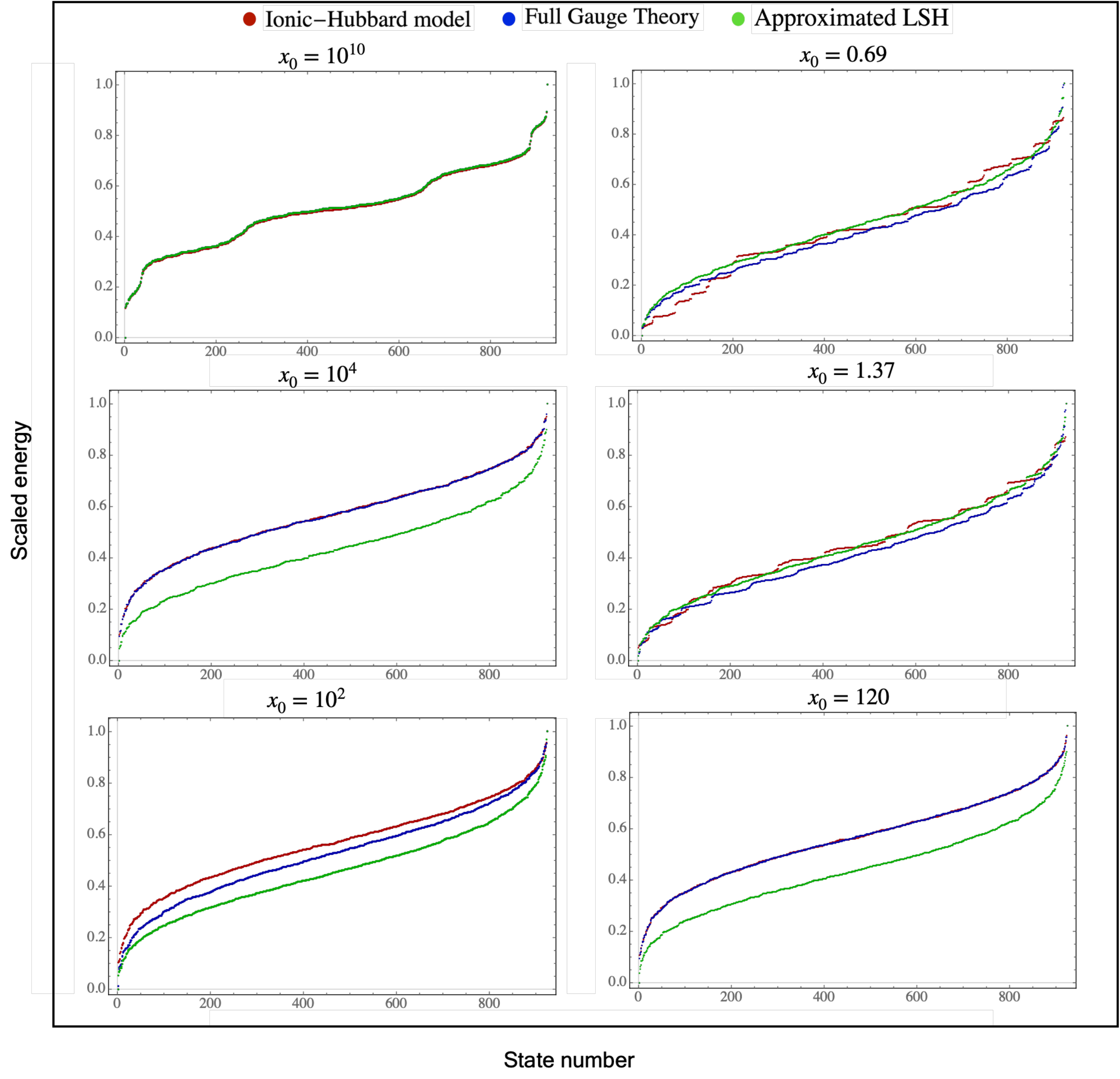}
\caption{Spectrum of the Ionic Hubbard model, Full SU(2) gauge theory (KS or LSH) Hamiltonian (without any approximation) and  the Weak coupling approximated LSH calculated by exact diagonalization for 6 site system and scaled to fit between 0 to 1.  The left panel shows the spectrum in weak coupling regime of gauge theory with parameters as per (\ref{wc_parameters}) for different values of $p$ as discussed in subsection \ref{sec:ws-scaling}. The right panel shows spectrum obtained for the string coupling analysis, for $V_0=17.5, 8.75 ~\& ~ 0.1$ respectively. The spectrum demonstrates that the intermediate coupling regime is better accessible by strong coupling analysis if smaller values of $V_0$ becomes experimentally feasible. We propose to quantum simulate strong coupling spectrum within a mean field approximation and at bulk limit, whereas the plots are only for small lattices and hence showing magnified deviation of the mean field spectrum from that of the full gauge theory. The approximated LSH is only valid in weak coupling regime and matches with full gauge theory for $p\gg1$.}
\label{fig:spectrum_Match}
\end{figure*}
\textbf{In weak coupling regime: }
We aim to quantum simulate gauge theory Hamiltonian, with the values of dimensionless parameters given by:
$$ x_0=10^{-10} ~~\& ~~ m/g=1.6\times 10^{-10} $$
acting on the LSH Hilbert space characterized by $$ n_l=5\times 10^{5}$$ at all sites and correspond to to $p=5,p'=0$ in (\ref{nlp}-\ref{pp'}). The fermionic (string) configurations remain completely dynamical as $n_i,n_o$ can take all possible values at sites $0,1,2,...2N$. Following (\ref{wcfix1}-\ref{wcfix2}) we obtain 
the parameters of the atomic Hamiltonian to be fixed at:
\bea
V_0=8.75t ~,~ V'=1.6t ~,~ u=0.1t  \label{wc_parameters}.
\eea
Note that, we have chosen a feasible but small value of the parameter $u$. Smaller and smaller values of $u$ will enable to mimic the dynamics of gauge theory more accurately as we take $p\gg1$.
We perform exact diagonalization for both the Hamiltonians with a small number of sites, that is doable on a PC. Our scheme, being completely scalable, the agreement in spectrum as in Fig. \ref{fig:spectrum_Match} holds true for any size of lattice as per experimental capabilities.

\textbf{In strong coupling regime:} 
We aim to quantum simulate gauge theory Hamiltonian (\ref{scaledHsc}), with the values of dimensionless parameters given by:
$$ x_0=0.69 ~~\& ~~ m/g=1.6. $$
This Hamiltonian acts on LSH Hilbert space characterized by $ l_i=6$ as in (\ref{nl-li}) , while $n_i,n_o$ can take all possible values at sites $0,1,2,...2N$.  Following (\ref{scfix1}-\ref{scfix2}), the mimicking atomic system is defined by parameters:
\bea
\label{sc_exp}
V_0=17.5t ~,~ V'=1.6t ~,~ u= 5.47t.
\eea
Likewise weak coupling case, we also perform exact diagonalization for this case to compare and obtain the spectrum as in Fig. \ref{fig:spectrum_Match}. Note that, from our analysis we only expect exact match of spectrum in $N\rightarrow \infty$ limit, that is beyond scope of exact diagonalization and is not reported here. Performing numerical calculations for a longer lattice is beyond scope of exact diagonalization, but can be performed using state of the art tensor network technique and that study would establish proper benchmark for the scheme in strong coupling regime. However, tensor network can only calculate a particular (low energy) sector of the theory with accuracy and quantum simulation is expected to outperform the same. 

\begin{figure*}[t]
\includegraphics[width=0.99\textwidth]{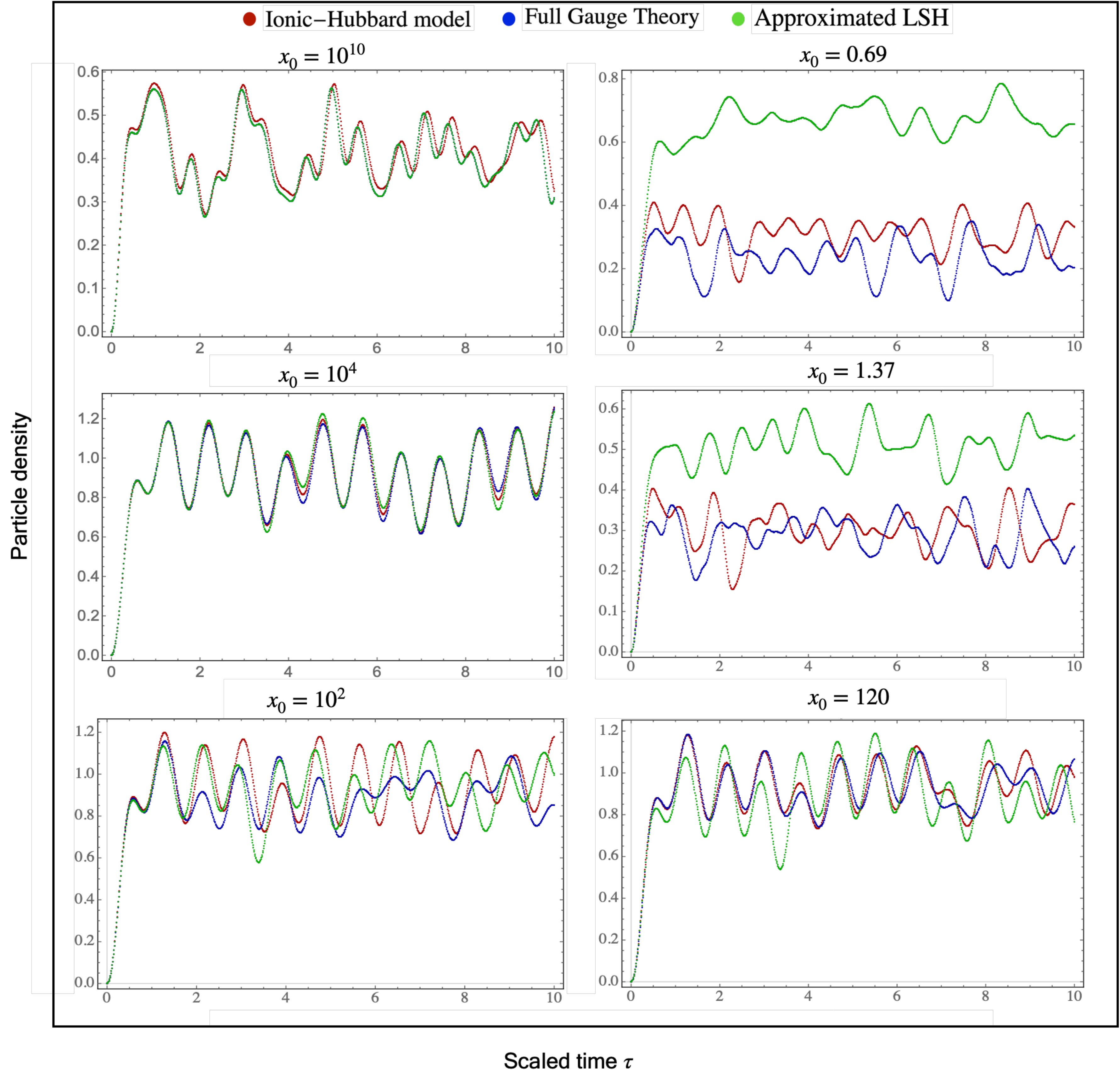}
\caption{Simulated particle density dynamics, corresponding to the cartoon of Fig. \ref{fig:dynamics_cartoon}(a) is plotted against a scaled time $\tau$. The parameters are identical to that used for spectrum analysis in Fig. \ref{fig:spectrum_Match}. The simulated dynamics is almost exact to that of the full gauge theory for weak coupling limit. The mismatch between full gauge theory dynamics and Hubbard model dynamics in strong coupling regime is expected to get minimized at bulk limit. The approximated LSH is only valid in weak coupling regime and matches with full gauge theory for $p\gg1$ as demonstrated in spectrum analysis as well. For the right panel, the simulated dynamics is matching better with the full gauge theory dynamics than that of the approximated Hamiltonian. This is because the tuned self interaction of atomic Hamiltonian takes care of a significant error that exists in the approximated Hamiltonian.}
\label{fig:dynamics_match}
\end{figure*}

However, even with limited computational resources, we make the following observations:
\begin{itemize}
    \item It appears from (\ref{scfix1}-\ref{scfix2}) that, by increasing $V_0/t$ in the atomic system, one would be able to access smaller and smaller values of the gauge theory parameter $x_0$. However, the consequence is that, in order to mimic exact strong coupling dynamics,  $u/t$ has to be increased as well.  
    \begin{figure}[h]
\includegraphics[width=0.48\textwidth]{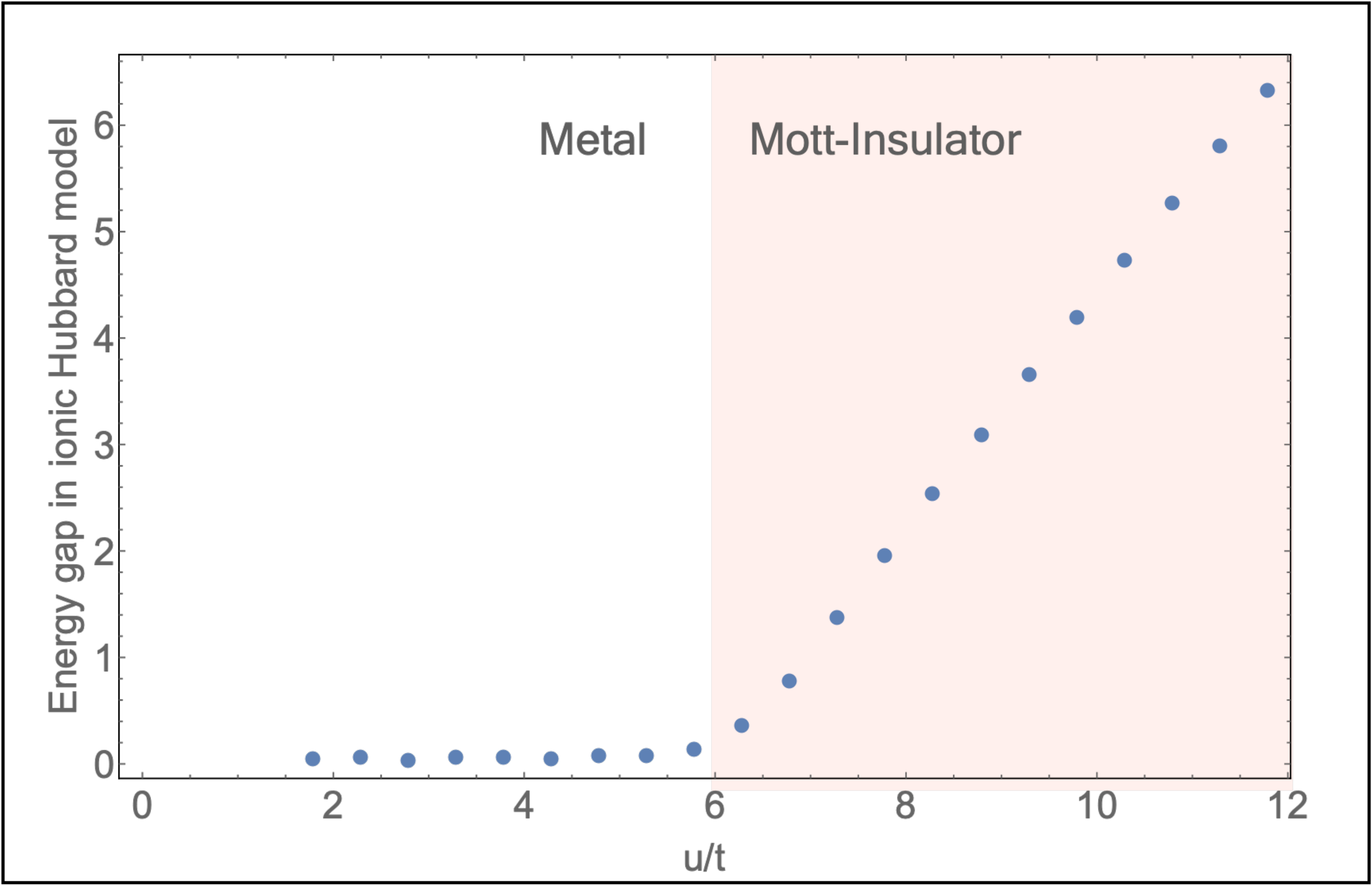}
\caption{A Quantum phase transition is observed with the ionic Hubbard model at a particular value of $u/t$, beyond which the spectrum becomes gapped and hence the Hubbard model can no longer mimic the dynamics of gauge theory. This particular plot is obtained with the parameters of Hubbard model given in (\ref{sc_exp}) except varying $u/t$. Choosing a larger value of $V_0/t$ corresponds to smaller value of $x_0$  via (\ref{fixV0}), but following (\ref{scfix2}) it will always be in the Mott Insulating phase.}
\label{fig:PT}
\end{figure}
    \item With an increasing $u/t$ (even for a fixed value of $V_0/t$), gaps are introduced in the  atomic spectrum  as the atomic system experiences a quantum phase transition (see Fig. \ref{fig:PT}) and enters into Mott Insulator phase \cite{imada1998metal}. Then the system can no longer  mimic dynamics of gauge theory as there is no such quantum phase transition in the gauge theory spectrum. Hence, this quantum simulation scheme is not suitable for $x_0\rightarrow 0$.
    \item Instead, if one can arrange the experimental set-up to fix $V_0/t$ at a smaller value, the atomic system simulates the intermediate coupling regime of the full gauge theory reliably. We illustrate such an agreement for $V_0/t=0.1 ~(x_0=120)$ in Fig \ref{fig:spectrum_Match}. 
\end{itemize}

\subsection{Simulated Dynamics:}

One important dynamical phenomenon to observe in real time dynamics in gauge theory is the dynamics of pair production and string breaking as illustrated via cartoon in Fig. \ref{fig:dynamics_cartoon}. We consider preparing the system in  a state in which all even sites are completely empty (no particle) and all odd sites are completely filled (no antiparticle). The real time Hamiltonian evolution of the atomic system involves  atoms hopping from one site to another, simulating the event of pair creation and particle number dynamics of gauge theory.  Within LSH framework, for the no particle-no antiparticle state  $|\Psi_0\rangle$ on a 1d lattice of $N$ staggered sites, we define the following quantity to describe particle density,
\bea
\rho(\tau) = 1+ \frac{1}{N}\langle \Psi_0|\hat{\mathcal U}^{\dagger}(\tau)\hat{\mathcal O}\hat{ \mathcal U}(\tau)|\Psi_0\rangle
\label{Ntgauge}
\eea
where, $\hat{\mathcal O}=\sum_{j}\left((-1)^j (\hat n_i(j)+\hat n_o(j))\right)$ and $\mathcal U(\tau_{\mathrm{gauge}})$ is defined in (\ref{unitarysc}).

The simulated dynamics in Hubbard model is measured by the observable  $P$ , as defined in (\ref{defP}). Its connection with the particle number dynamics of gauge theory can be obtained by looking at the parameter $1+P$. In Fig. \ref{fig:dynamics_match} we plot the quantities against a scaled time $\tau= \tau_{\mbox{atomic}}=2a\tau_{\mathrm{gauge}}$ following (\ref{scaledtime_gauge}). 

As done in the spectrum analysis, we consider the same parameter values  for calculating pair-production and string breaking dynamics as well. From the simulated dynamics we can conclude the following :
\begin{itemize} 
\item The proposed simulation scheme simulates the dynamics of weak coupling gauge theory perfectly and that is evident even from the numerical analysis using a small system. Here the particle density dynamics resulting from i) full gauge theory, ii) the approximated LSH theory and iii) the atomic Hamiltonian all agree very well.
\item The difference between the actual dynamics due to the original Hamiltonian and the dynamics due to the approximated Hamiltonian is quite pronounced in the intermediate coupling/ strong coupling regimes.  However, by adjusting the on-site interaction parameter, it was possible to recover the correction in the electric energy term (\ref{corr_HE}) substantially, and hence the ionic-Hubbard dynamics is now closer to the dynamics of the full gauge theory, when compared to the same with the approximated LSH formulation.
\item The discrepancy that still exists in the intermediate/ strong coupling regimes will surely get reduced if one can simulate using a long enough lattice, such that in the statistical limit, one can really recover the correction in electric energy term (\ref{corr_HE}) in full by choosing the atomic self-interaction accordingly. Considering that we used a small lattice (6 site system) for our numerical simulation and yet managed to observe a good agreement, it is extremely likely that in an actual experiment (or, tensor network calculation) involving a large number of lattice sites, the error will be insignificant. 
\end{itemize} 
It is discussed in Sec. \ref{sec:exp}, how one can measure the dynamics in an actual experiment. However, the actual time measured in ms during the experiment is related to the scaled times as:
\bea
\tau_{\mbox{exp}} &=& \frac{\hbar \tau_{\mbox{atomic}}}{t} \equiv \frac{\tau_{\mbox{atomic}}}{1.5716}~ \mbox{ms}\\
\Rightarrow & \equiv& \frac{2a\tau_{\mbox{gauge}}}{1.5716} \mbox{ms}.
\eea
Thus, for different values of lattice spacing, the same experiment would simulate real time dynamics of gauge theory happening in different smaller time scale.

\section{Discussions and Future Directions}
\label{sec:conclusion}
\noindent
This paper presents the very first practically implementable quantum simulation proposal for simulating SU(2) lattice gauge theory in $(1+1)$-d, that specifically simulates the spectrum and dynamics of gauge theory in weak coupling regime as well as intermediate coupling regime for a large lattice with good accuracy. Experimental implementation of this particular scheme will demonstrate why quantum simulators can be a very effective tool to study different aspects of gauge theories. 

The proposal is completely scalable that accesses different regimes of gauge theory (with a varying degree of accuracy) and quantum simulate different symmetry sectors. A suitable scaling scheme presented in this paper enables one to model different regimes of LGT with a single experimental set up, just by tuning the controllable experimental parameters.  For example, the weak and strong coupling limits of gauge theory is accessed by taking $u/t$ to $0$ and $u< u_c$ respectively in the atomic system, where $u_c$ is a quantum critical point beyond which the atomic system enters into a Mott insulating phase as observed in this particular study with small lattice (see Fig. \ref{fig:PT}). The only requirement here is that the system requires to remain in the same paramagnetic phase throughout the course of its dynamics, so that in a bulk limit, all the allowed states are equally probable at half-filling.

Future works will address the issue of going beyond mean field approximation by simulating dynamical gauge fields that exists beyond 1 spatial dimension. The LSH formalism for gauge theories in higher dimensions should be equally useful in constructing atomic quantum simulators for the same. Specifically, within LSH framework, the matter gauge coupling remains the same as in 1d in any higher dimension, including the feature of non-dynamic loop degrees of freedom at matter sites \cite{Raychowdhury:2018osk,Raychowdhury:2018tfj,Raychowdhury:2019iki}. Hence we expect the present proposal to remain as a useful building block for higher dimensional quantum simulators as well. Work is in progress in these directions and will be reported elsewhere. The present scheme can also be generalized for gauge group SU(3) upon generalization of LSH formalism for SU(3) gauge theory and that will build a concrete step towards quantum simulating QCD.


\section*{ACKNOWLEDGEMENT}

We would like to thank Zohreh Davoudi and Rudranil Basu for useful discussions and also for careful reading of the manuscript and helpful comments. R.D. would like to acknowledge support from the Department of Science and Technology, Government of India in the form of an Inspire Faculty Award (Grant No. 04/2014/002342). I. R. is supported by the U.S. Department of Energy
(DOE), Office of Science, Office of Advanced Scientific
Computing Research (ASCR) Quantum Computing Application Teams (QCAT) program, under fieldwork Proposal No. ERKJ347. 
\bibliography{bibi.bib}

\begin{thebibliography}{74}%
\makeatletter
\providecommand \@ifxundefined [1]{%
 \@ifx{#1\undefined}
}%
\providecommand \@ifnum [1]{%
 \ifnum #1\expandafter \@firstoftwo
 \else \expandafter \@secondoftwo
 \fi
}%
\providecommand \@ifx [1]{%
 \ifx #1\expandafter \@firstoftwo
 \else \expandafter \@secondoftwo
 \fi
}%
\providecommand \natexlab [1]{#1}%
\providecommand \enquote  [1]{``#1''}%
\providecommand \bibnamefont  [1]{#1}%
\providecommand \bibfnamefont [1]{#1}%
\providecommand \citenamefont [1]{#1}%
\providecommand \href@noop [0]{\@secondoftwo}%
\providecommand \href [0]{\begingroup \@sanitize@url \@href}%
\providecommand \@href[1]{\@@startlink{#1}\@@href}%
\providecommand \@@href[1]{\endgroup#1\@@endlink}%
\providecommand \@sanitize@url [0]{\catcode `\\12\catcode `\$12\catcode
  `\&12\catcode `\#12\catcode `\^12\catcode `\_12\catcode `\%12\relax}%
\providecommand \@@startlink[1]{}%
\providecommand \@@endlink[0]{}%
\providecommand \url  [0]{\begingroup\@sanitize@url \@url }%
\providecommand \@url [1]{\endgroup\@href {#1}{\urlprefix }}%
\providecommand \urlprefix  [0]{URL }%
\providecommand \Eprint [0]{\href }%
\providecommand \doibase [0]{http://dx.doi.org/}%
\providecommand \selectlanguage [0]{\@gobble}%
\providecommand \bibinfo  [0]{\@secondoftwo}%
\providecommand \bibfield  [0]{\@secondoftwo}%
\providecommand \translation [1]{[#1]}%
\providecommand \BibitemOpen [0]{}%
\providecommand \bibitemStop [0]{}%
\providecommand \bibitemNoStop [0]{.\EOS\space}%
\providecommand \EOS [0]{\spacefactor3000\relax}%
\providecommand \BibitemShut  [1]{\csname bibitem#1\endcsname}%
\let\auto@bib@innerbib\@empty
\bibitem [{\citenamefont {Wilson}(1974)}]{Wilson:1974sk}%
  \BibitemOpen
  \bibfield  {author} {\bibinfo {author} {\bibfnamefont {Kenneth~G.}\
  \bibnamefont {Wilson}},\ }\bibfield  {title} {\enquote {\bibinfo {title}
  {Confinement of quarks},}\ }\href {\doibase 10.1103/PhysRevD.10.2445}
  {\bibfield  {journal} {\bibinfo  {journal} {Phys. Rev. D}\ }\textbf {\bibinfo
  {volume} {10}},\ \bibinfo {pages} {2445--2459} (\bibinfo {year}
  {1974})}\BibitemShut {NoStop}%
\bibitem [{\citenamefont {Creutz}\ \emph {et~al.}(1983)\citenamefont {Creutz},
  \citenamefont {Jacobs},\ and\ \citenamefont {Rebbi}}]{CREUTZ1983201}%
  \BibitemOpen
  \bibfield  {author} {\bibinfo {author} {\bibfnamefont {Michael}\ \bibnamefont
  {Creutz}}, \bibinfo {author} {\bibfnamefont {Laurence}\ \bibnamefont
  {Jacobs}}, \ and\ \bibinfo {author} {\bibfnamefont {Claudio}\ \bibnamefont
  {Rebbi}},\ }\bibfield  {title} {\enquote {\bibinfo {title} {Monte carlo
  computations in lattice gauge theories},}\ }\href {\doibase
  https://doi.org/10.1016/0370-1573(83)90016-9} {\bibfield  {journal} {\bibinfo
   {journal} {Physics Reports}\ }\textbf {\bibinfo {volume} {95}},\ \bibinfo
  {pages} {201 -- 282} (\bibinfo {year} {1983})}\BibitemShut {NoStop}%
\bibitem [{\citenamefont {Jo{\'o}}\ \emph {et~al.}(2019)\citenamefont
  {Jo{\'o}}, \citenamefont {Jung}, \citenamefont {Christ}, \citenamefont
  {Detmold}, \citenamefont {Edwards}, \citenamefont {Savage},\ and\
  \citenamefont {Shanahan}}]{joo2019status}%
  \BibitemOpen
  \bibfield  {author} {\bibinfo {author} {\bibfnamefont {B{\'a}lint}\
  \bibnamefont {Jo{\'o}}}, \bibinfo {author} {\bibfnamefont {Chulwoo}\
  \bibnamefont {Jung}}, \bibinfo {author} {\bibfnamefont {Norman~H}\
  \bibnamefont {Christ}}, \bibinfo {author} {\bibfnamefont {William}\
  \bibnamefont {Detmold}}, \bibinfo {author} {\bibfnamefont {Robert~G}\
  \bibnamefont {Edwards}}, \bibinfo {author} {\bibfnamefont {Martin}\
  \bibnamefont {Savage}}, \ and\ \bibinfo {author} {\bibfnamefont {Phiala}\
  \bibnamefont {Shanahan}},\ }\bibfield  {title} {\enquote {\bibinfo {title}
  {Status and future perspectives for lattice gauge theory calculations to the
  exascale and beyond},}\ }\href@noop {} {\bibfield  {journal} {\bibinfo
  {journal} {The European Physical Journal A}\ }\textbf {\bibinfo {volume}
  {55}},\ \bibinfo {pages} {199} (\bibinfo {year} {2019})}\BibitemShut
  {NoStop}%
\bibitem [{\citenamefont {De~Forcrand}(2010)}]{de2010simulating}%
  \BibitemOpen
  \bibfield  {author} {\bibinfo {author} {\bibfnamefont {Philippe}\
  \bibnamefont {De~Forcrand}},\ }\bibfield  {title} {\enquote {\bibinfo {title}
  {Simulating qcd at finite density},}\ }\href@noop {} {\bibfield  {journal}
  {\bibinfo  {journal} {arXiv preprint arXiv:1005.0539}\ } (\bibinfo {year}
  {2010})}\BibitemShut {NoStop}%
\bibitem [{\citenamefont {Wiese}(2013)}]{wiese2013ultracold}%
  \BibitemOpen
  \bibfield  {author} {\bibinfo {author} {\bibfnamefont {U-J}\ \bibnamefont
  {Wiese}},\ }\bibfield  {title} {\enquote {\bibinfo {title} {Ultracold quantum
  gases and lattice systems: quantum simulation of lattice gauge theories},}\
  }\href@noop {} {\bibfield  {journal} {\bibinfo  {journal} {Annalen der
  Physik}\ }\textbf {\bibinfo {volume} {525}},\ \bibinfo {pages} {777--796}
  (\bibinfo {year} {2013})}\BibitemShut {NoStop}%
\bibitem [{\citenamefont {Feynman}(1982)}]{feynman1982simulating}%
  \BibitemOpen
  \bibfield  {author} {\bibinfo {author} {\bibfnamefont {Richard~P}\
  \bibnamefont {Feynman}},\ }\bibfield  {title} {\enquote {\bibinfo {title}
  {Simulating physics with computers},}\ }\href@noop {} {\bibfield  {journal}
  {\bibinfo  {journal} {Int. J. Theor. Phys}\ }\textbf {\bibinfo {volume} {21}}
  (\bibinfo {year} {1982})}\BibitemShut {NoStop}%
\bibitem [{\citenamefont {Bloch}\ \emph {et~al.}(2012)\citenamefont {Bloch},
  \citenamefont {Dalibard},\ and\ \citenamefont
  {Nascimbene}}]{bloch2012quantum}%
  \BibitemOpen
  \bibfield  {author} {\bibinfo {author} {\bibfnamefont {Immanuel}\
  \bibnamefont {Bloch}}, \bibinfo {author} {\bibfnamefont {Jean}\ \bibnamefont
  {Dalibard}}, \ and\ \bibinfo {author} {\bibfnamefont {Sylvain}\ \bibnamefont
  {Nascimbene}},\ }\bibfield  {title} {\enquote {\bibinfo {title} {Quantum
  simulations with ultracold quantum gases},}\ }\href@noop {} {\bibfield
  {journal} {\bibinfo  {journal} {Nature Physics}\ }\textbf {\bibinfo {volume}
  {8}},\ \bibinfo {pages} {267--276} (\bibinfo {year} {2012})}\BibitemShut
  {NoStop}%
\bibitem [{\citenamefont {Blatt}\ and\ \citenamefont
  {Roos}(2012)}]{blatt2012quantum}%
  \BibitemOpen
  \bibfield  {author} {\bibinfo {author} {\bibfnamefont {Rainer}\ \bibnamefont
  {Blatt}}\ and\ \bibinfo {author} {\bibfnamefont {Christian~F}\ \bibnamefont
  {Roos}},\ }\bibfield  {title} {\enquote {\bibinfo {title} {Quantum
  simulations with trapped ions},}\ }\href@noop {} {\bibfield  {journal}
  {\bibinfo  {journal} {Nature Physics}\ }\textbf {\bibinfo {volume} {8}},\
  \bibinfo {pages} {277--284} (\bibinfo {year} {2012})}\BibitemShut {NoStop}%
\bibitem [{\citenamefont {Zohar}\ and\ \citenamefont
  {Reznik}(2011)}]{zohar2011confinement}%
  \BibitemOpen
  \bibfield  {author} {\bibinfo {author} {\bibfnamefont {Erez}\ \bibnamefont
  {Zohar}}\ and\ \bibinfo {author} {\bibfnamefont {Benni}\ \bibnamefont
  {Reznik}},\ }\bibfield  {title} {\enquote {\bibinfo {title} {Confinement and
  lattice quantum-electrodynamic electric flux tubes simulated with ultracold
  atoms},}\ }\href@noop {} {\bibfield  {journal} {\bibinfo  {journal} {Physical
  review letters}\ }\textbf {\bibinfo {volume} {107}},\ \bibinfo {pages}
  {275301} (\bibinfo {year} {2011})}\BibitemShut {NoStop}%
\bibitem [{\citenamefont {Zohar}\ \emph
  {et~al.}(2013{\natexlab{a}})\citenamefont {Zohar}, \citenamefont {Cirac},\
  and\ \citenamefont {Reznik}}]{Zohar:2012xf}%
  \BibitemOpen
  \bibfield  {author} {\bibinfo {author} {\bibfnamefont {Erez}\ \bibnamefont
  {Zohar}}, \bibinfo {author} {\bibfnamefont {J.Ignacio}\ \bibnamefont
  {Cirac}}, \ and\ \bibinfo {author} {\bibfnamefont {Benni}\ \bibnamefont
  {Reznik}},\ }\bibfield  {title} {\enquote {\bibinfo {title} {{Cold-Atom
  Quantum Simulator for SU(2) Yang-Mills Lattice Gauge Theory}},}\ }\href
  {\doibase 10.1103/PhysRevLett.110.125304} {\bibfield  {journal} {\bibinfo
  {journal} {Phys. Rev. Lett.}\ }\textbf {\bibinfo {volume} {110}},\ \bibinfo
  {pages} {125304} (\bibinfo {year} {2013}{\natexlab{a}})},\ \Eprint
  {http://arxiv.org/abs/1211.2241} {arXiv:1211.2241 [quant-ph]} \BibitemShut
  {NoStop}%
\bibitem [{\citenamefont {Zohar}\ \emph
  {et~al.}(2013{\natexlab{b}})\citenamefont {Zohar}, \citenamefont {Cirac},\
  and\ \citenamefont {Reznik}}]{Zohar:2013zla}%
  \BibitemOpen
  \bibfield  {author} {\bibinfo {author} {\bibfnamefont {Erez}\ \bibnamefont
  {Zohar}}, \bibinfo {author} {\bibfnamefont {J.~Ignacio}\ \bibnamefont
  {Cirac}}, \ and\ \bibinfo {author} {\bibfnamefont {Benni}\ \bibnamefont
  {Reznik}},\ }\bibfield  {title} {\enquote {\bibinfo {title} {{Quantum
  simulations of gauge theories with ultracold atoms: local gauge invariance
  from angular momentum conservation}},}\ }\href {\doibase
  10.1103/PhysRevA.88.023617} {\bibfield  {journal} {\bibinfo  {journal} {Phys.
  Rev. A}\ }\textbf {\bibinfo {volume} {88}},\ \bibinfo {pages} {023617}
  (\bibinfo {year} {2013}{\natexlab{b}})},\ \Eprint
  {http://arxiv.org/abs/1303.5040} {arXiv:1303.5040 [quant-ph]} \BibitemShut
  {NoStop}%
\bibitem [{\citenamefont {Zohar}\ \emph {et~al.}(2015)\citenamefont {Zohar},
  \citenamefont {Cirac},\ and\ \citenamefont {Reznik}}]{Zohar:2015hwa}%
  \BibitemOpen
  \bibfield  {author} {\bibinfo {author} {\bibfnamefont {Erez}\ \bibnamefont
  {Zohar}}, \bibinfo {author} {\bibfnamefont {J~Ignacio}\ \bibnamefont
  {Cirac}}, \ and\ \bibinfo {author} {\bibfnamefont {Benni}\ \bibnamefont
  {Reznik}},\ }\bibfield  {title} {\enquote {\bibinfo {title} {Quantum
  simulations of lattice gauge theories using ultracold atoms in optical
  lattices},}\ }\href@noop {} {\bibfield  {journal} {\bibinfo  {journal}
  {Reports on Progress in Physics}\ }\textbf {\bibinfo {volume} {79}},\
  \bibinfo {pages} {014401} (\bibinfo {year} {2015})}\BibitemShut {NoStop}%
\bibitem [{\citenamefont {Banerjee}\ \emph {et~al.}(2012)\citenamefont
  {Banerjee}, \citenamefont {Dalmonte}, \citenamefont {Muller}, \citenamefont
  {Rico}, \citenamefont {Stebler}, \citenamefont {Wiese},\ and\ \citenamefont
  {Zoller}}]{Banerjee:2012pg}%
  \BibitemOpen
  \bibfield  {author} {\bibinfo {author} {\bibfnamefont {D.}~\bibnamefont
  {Banerjee}}, \bibinfo {author} {\bibfnamefont {M.}~\bibnamefont {Dalmonte}},
  \bibinfo {author} {\bibfnamefont {M.}~\bibnamefont {Muller}}, \bibinfo
  {author} {\bibfnamefont {E.}~\bibnamefont {Rico}}, \bibinfo {author}
  {\bibfnamefont {P.}~\bibnamefont {Stebler}}, \bibinfo {author} {\bibfnamefont
  {U.-J.}\ \bibnamefont {Wiese}}, \ and\ \bibinfo {author} {\bibfnamefont
  {P.}~\bibnamefont {Zoller}},\ }\bibfield  {title} {\enquote {\bibinfo {title}
  {{Atomic Quantum Simulation of Dynamical Gauge Fields coupled to Fermionic
  Matter: From String Breaking to Evolution after a Quench}},}\ }\href
  {\doibase 10.1103/PhysRevLett.109.175302} {\bibfield  {journal} {\bibinfo
  {journal} {Phys. Rev. Lett.}\ }\textbf {\bibinfo {volume} {109}},\ \bibinfo
  {pages} {175302} (\bibinfo {year} {2012})},\ \Eprint
  {http://arxiv.org/abs/1205.6366} {arXiv:1205.6366 [cond-mat.quant-gas]}
  \BibitemShut {NoStop}%
\bibitem [{\citenamefont {Banerjee}\ \emph {et~al.}(2013)\citenamefont
  {Banerjee}, \citenamefont {Bögli}, \citenamefont {Dalmonte}, \citenamefont
  {Rico}, \citenamefont {Stebler}, \citenamefont {Wiese},\ and\ \citenamefont
  {Zoller}}]{Banerjee:2012xg}%
  \BibitemOpen
  \bibfield  {author} {\bibinfo {author} {\bibfnamefont {D.}~\bibnamefont
  {Banerjee}}, \bibinfo {author} {\bibfnamefont {M.}~\bibnamefont {Bögli}},
  \bibinfo {author} {\bibfnamefont {M.}~\bibnamefont {Dalmonte}}, \bibinfo
  {author} {\bibfnamefont {E.}~\bibnamefont {Rico}}, \bibinfo {author}
  {\bibfnamefont {P.}~\bibnamefont {Stebler}}, \bibinfo {author} {\bibfnamefont
  {U.-J.}\ \bibnamefont {Wiese}}, \ and\ \bibinfo {author} {\bibfnamefont
  {P.}~\bibnamefont {Zoller}},\ }\bibfield  {title} {\enquote {\bibinfo {title}
  {{Atomic Quantum Simulation of U(N) and SU(N) Non-Abelian Lattice Gauge
  Theories}},}\ }\href {\doibase 10.1103/PhysRevLett.110.125303} {\bibfield
  {journal} {\bibinfo  {journal} {Phys. Rev. Lett.}\ }\textbf {\bibinfo
  {volume} {110}},\ \bibinfo {pages} {125303} (\bibinfo {year} {2013})},\
  \Eprint {http://arxiv.org/abs/1211.2242} {arXiv:1211.2242
  [cond-mat.quant-gas]} \BibitemShut {NoStop}%
\bibitem [{\citenamefont {Stannigel}\ \emph {et~al.}(2014)\citenamefont
  {Stannigel}, \citenamefont {Hauke}, \citenamefont {Marcos}, \citenamefont
  {Hafezi}, \citenamefont {Diehl}, \citenamefont {Dalmonte},\ and\
  \citenamefont {Zoller}}]{stannigel2014constrained}%
  \BibitemOpen
  \bibfield  {author} {\bibinfo {author} {\bibfnamefont {K}~\bibnamefont
  {Stannigel}}, \bibinfo {author} {\bibfnamefont {P}~\bibnamefont {Hauke}},
  \bibinfo {author} {\bibfnamefont {D}~\bibnamefont {Marcos}}, \bibinfo
  {author} {\bibfnamefont {M}~\bibnamefont {Hafezi}}, \bibinfo {author}
  {\bibfnamefont {S}~\bibnamefont {Diehl}}, \bibinfo {author} {\bibfnamefont
  {M}~\bibnamefont {Dalmonte}}, \ and\ \bibinfo {author} {\bibfnamefont
  {P}~\bibnamefont {Zoller}},\ }\bibfield  {title} {\enquote {\bibinfo {title}
  {Constrained dynamics via the zeno effect in quantum simulation: Implementing
  non-abelian lattice gauge theories with cold atoms},}\ }\href@noop {}
  {\bibfield  {journal} {\bibinfo  {journal} {Physical review letters}\
  }\textbf {\bibinfo {volume} {112}},\ \bibinfo {pages} {120406} (\bibinfo
  {year} {2014})}\BibitemShut {NoStop}%
\bibitem [{\citenamefont {Gonz\'alez-Cuadra}\ \emph {et~al.}(2017)\citenamefont
  {Gonz\'alez-Cuadra}, \citenamefont {Zohar},\ and\ \citenamefont
  {Cirac}}]{Gonzalez-Cuadra:2017lvz}%
  \BibitemOpen
  \bibfield  {author} {\bibinfo {author} {\bibfnamefont {Daniel}\ \bibnamefont
  {Gonz\'alez-Cuadra}}, \bibinfo {author} {\bibfnamefont {Erez}\ \bibnamefont
  {Zohar}}, \ and\ \bibinfo {author} {\bibfnamefont {J.~Ignacio}\ \bibnamefont
  {Cirac}},\ }\bibfield  {title} {\enquote {\bibinfo {title} {{Quantum
  Simulation of the Abelian-Higgs Lattice Gauge Theory with Ultracold
  Atoms}},}\ }\href {\doibase 10.1088/1367-2630/aa6f37} {\bibfield  {journal}
  {\bibinfo  {journal} {New J. Phys.}\ }\textbf {\bibinfo {volume} {19}},\
  \bibinfo {pages} {063038} (\bibinfo {year} {2017})},\ \Eprint
  {http://arxiv.org/abs/1702.05492} {arXiv:1702.05492 [quant-ph]} \BibitemShut
  {NoStop}%
\bibitem [{\citenamefont {Tagliacozzo}\ \emph {et~al.}(2013)\citenamefont
  {Tagliacozzo}, \citenamefont {Celi}, \citenamefont {Orland}, \citenamefont
  {Mitchell},\ and\ \citenamefont {Lewenstein}}]{tagliacozzo2013simulation}%
  \BibitemOpen
  \bibfield  {author} {\bibinfo {author} {\bibfnamefont {L}~\bibnamefont
  {Tagliacozzo}}, \bibinfo {author} {\bibfnamefont {A}~\bibnamefont {Celi}},
  \bibinfo {author} {\bibfnamefont {P}~\bibnamefont {Orland}}, \bibinfo
  {author} {\bibfnamefont {MW}~\bibnamefont {Mitchell}}, \ and\ \bibinfo
  {author} {\bibfnamefont {M}~\bibnamefont {Lewenstein}},\ }\bibfield  {title}
  {\enquote {\bibinfo {title} {Simulation of non-abelian gauge theories with
  optical lattices},}\ }\href@noop {} {\bibfield  {journal} {\bibinfo
  {journal} {Nature communications}\ }\textbf {\bibinfo {volume} {4}},\
  \bibinfo {pages} {1--8} (\bibinfo {year} {2013})}\BibitemShut {NoStop}%
\bibitem [{\citenamefont {Kasper}\ \emph {et~al.}(2016)\citenamefont {Kasper},
  \citenamefont {Hebenstreit}, \citenamefont {Oberthaler},\ and\ \citenamefont
  {Berges}}]{kasper2016schwinger}%
  \BibitemOpen
  \bibfield  {author} {\bibinfo {author} {\bibfnamefont {V}~\bibnamefont
  {Kasper}}, \bibinfo {author} {\bibfnamefont {Florian}\ \bibnamefont
  {Hebenstreit}}, \bibinfo {author} {\bibfnamefont {MK}~\bibnamefont
  {Oberthaler}}, \ and\ \bibinfo {author} {\bibfnamefont {J}~\bibnamefont
  {Berges}},\ }\bibfield  {title} {\enquote {\bibinfo {title} {Schwinger pair
  production with ultracold atoms},}\ }\href@noop {} {\bibfield  {journal}
  {\bibinfo  {journal} {Physics Letters B}\ }\textbf {\bibinfo {volume}
  {760}},\ \bibinfo {pages} {742--746} (\bibinfo {year} {2016})}\BibitemShut
  {NoStop}%
\bibitem [{\citenamefont {Schweizer}\ \emph {et~al.}(2019)\citenamefont
  {Schweizer}, \citenamefont {Grusdt}, \citenamefont {Berngruber},
  \citenamefont {Barbiero}, \citenamefont {Demler}, \citenamefont {Goldman},
  \citenamefont {Bloch},\ and\ \citenamefont
  {Aidelsburger}}]{schweizer2019floquet}%
  \BibitemOpen
  \bibfield  {author} {\bibinfo {author} {\bibfnamefont {Christian}\
  \bibnamefont {Schweizer}}, \bibinfo {author} {\bibfnamefont {Fabian}\
  \bibnamefont {Grusdt}}, \bibinfo {author} {\bibfnamefont {Moritz}\
  \bibnamefont {Berngruber}}, \bibinfo {author} {\bibfnamefont {Luca}\
  \bibnamefont {Barbiero}}, \bibinfo {author} {\bibfnamefont {Eugene}\
  \bibnamefont {Demler}}, \bibinfo {author} {\bibfnamefont {Nathan}\
  \bibnamefont {Goldman}}, \bibinfo {author} {\bibfnamefont {Immanuel}\
  \bibnamefont {Bloch}}, \ and\ \bibinfo {author} {\bibfnamefont {Monika}\
  \bibnamefont {Aidelsburger}},\ }\bibfield  {title} {\enquote {\bibinfo
  {title} {Floquet approach to z2 lattice gauge theories with ultracold atoms
  in optical lattices},}\ }\href@noop {} {\bibfield  {journal} {\bibinfo
  {journal} {Nature Physics}\ }\textbf {\bibinfo {volume} {15}},\ \bibinfo
  {pages} {1168--1173} (\bibinfo {year} {2019})}\BibitemShut {NoStop}%
\bibitem [{\citenamefont {Mil}\ \emph {et~al.}(2020)\citenamefont {Mil},
  \citenamefont {Zache}, \citenamefont {Hegde}, \citenamefont {Xia},
  \citenamefont {Bhatt}, \citenamefont {Oberthaler}, \citenamefont {Hauke},
  \citenamefont {Berges},\ and\ \citenamefont
  {Jendrzejewski}}]{mil2020scalable}%
  \BibitemOpen
  \bibfield  {author} {\bibinfo {author} {\bibfnamefont {Alexander}\
  \bibnamefont {Mil}}, \bibinfo {author} {\bibfnamefont {Torsten~V}\
  \bibnamefont {Zache}}, \bibinfo {author} {\bibfnamefont {Apoorva}\
  \bibnamefont {Hegde}}, \bibinfo {author} {\bibfnamefont {Andy}\ \bibnamefont
  {Xia}}, \bibinfo {author} {\bibfnamefont {Rohit~P}\ \bibnamefont {Bhatt}},
  \bibinfo {author} {\bibfnamefont {Markus~K}\ \bibnamefont {Oberthaler}},
  \bibinfo {author} {\bibfnamefont {Philipp}\ \bibnamefont {Hauke}}, \bibinfo
  {author} {\bibfnamefont {J{\"u}rgen}\ \bibnamefont {Berges}}, \ and\ \bibinfo
  {author} {\bibfnamefont {Fred}\ \bibnamefont {Jendrzejewski}},\ }\bibfield
  {title} {\enquote {\bibinfo {title} {A scalable realization of local u (1)
  gauge invariance in cold atomic mixtures},}\ }\href@noop {} {\bibfield
  {journal} {\bibinfo  {journal} {Science}\ }\textbf {\bibinfo {volume}
  {367}},\ \bibinfo {pages} {1128--1130} (\bibinfo {year} {2020})}\BibitemShut
  {NoStop}%
\bibitem [{\citenamefont {Yang}\ \emph
  {et~al.}(2020{\natexlab{a}})\citenamefont {Yang}, \citenamefont {Sun},
  \citenamefont {Ott}, \citenamefont {Wang}, \citenamefont {Zache},
  \citenamefont {Halimeh}, \citenamefont {Yuan}, \citenamefont {Hauke},\ and\
  \citenamefont {Pan}}]{yang2020observation}%
  \BibitemOpen
  \bibfield  {author} {\bibinfo {author} {\bibfnamefont {Bing}\ \bibnamefont
  {Yang}}, \bibinfo {author} {\bibfnamefont {Hui}\ \bibnamefont {Sun}},
  \bibinfo {author} {\bibfnamefont {Robert}\ \bibnamefont {Ott}}, \bibinfo
  {author} {\bibfnamefont {Han-Yi}\ \bibnamefont {Wang}}, \bibinfo {author}
  {\bibfnamefont {Torsten~V}\ \bibnamefont {Zache}}, \bibinfo {author}
  {\bibfnamefont {Jad~C}\ \bibnamefont {Halimeh}}, \bibinfo {author}
  {\bibfnamefont {Zhen-Sheng}\ \bibnamefont {Yuan}}, \bibinfo {author}
  {\bibfnamefont {Philipp}\ \bibnamefont {Hauke}}, \ and\ \bibinfo {author}
  {\bibfnamefont {Jian-Wei}\ \bibnamefont {Pan}},\ }\bibfield  {title}
  {\enquote {\bibinfo {title} {Observation of gauge invariance in a 71-site
  quantum simulator},}\ }\href@noop {} {\bibfield  {journal} {\bibinfo
  {journal} {arXiv preprint arXiv:2003.08945}\ } (\bibinfo {year}
  {2020}{\natexlab{a}})}\BibitemShut {NoStop}%
\bibitem [{\citenamefont {Davoudi}\ \emph
  {et~al.}(2020{\natexlab{a}})\citenamefont {Davoudi}, \citenamefont {Hafezi},
  \citenamefont {Monroe}, \citenamefont {Pagano}, \citenamefont {Seif},\ and\
  \citenamefont {Shaw}}]{davoudi2020towards}%
  \BibitemOpen
  \bibfield  {author} {\bibinfo {author} {\bibfnamefont {Zohreh}\ \bibnamefont
  {Davoudi}}, \bibinfo {author} {\bibfnamefont {Mohammad}\ \bibnamefont
  {Hafezi}}, \bibinfo {author} {\bibfnamefont {Christopher}\ \bibnamefont
  {Monroe}}, \bibinfo {author} {\bibfnamefont {Guido}\ \bibnamefont {Pagano}},
  \bibinfo {author} {\bibfnamefont {Alireza}\ \bibnamefont {Seif}}, \ and\
  \bibinfo {author} {\bibfnamefont {Andrew}\ \bibnamefont {Shaw}},\ }\bibfield
  {title} {\enquote {\bibinfo {title} {Towards analog quantum simulations of
  lattice gauge theories with trapped ions},}\ }\href@noop {} {\bibfield
  {journal} {\bibinfo  {journal} {Physical Review Research}\ }\textbf {\bibinfo
  {volume} {2}},\ \bibinfo {pages} {023015} (\bibinfo {year}
  {2020}{\natexlab{a}})}\BibitemShut {NoStop}%
\bibitem [{\citenamefont {Anderson}\ \emph {et~al.}(1995)\citenamefont
  {Anderson}, \citenamefont {Ensher}, \citenamefont {Matthews}, \citenamefont
  {Wieman},\ and\ \citenamefont {Cornell}}]{anderson1995observation}%
  \BibitemOpen
  \bibfield  {author} {\bibinfo {author} {\bibfnamefont {Mike~H}\ \bibnamefont
  {Anderson}}, \bibinfo {author} {\bibfnamefont {Jason~R}\ \bibnamefont
  {Ensher}}, \bibinfo {author} {\bibfnamefont {Michael~R}\ \bibnamefont
  {Matthews}}, \bibinfo {author} {\bibfnamefont {Carl~E}\ \bibnamefont
  {Wieman}}, \ and\ \bibinfo {author} {\bibfnamefont {Eric~A}\ \bibnamefont
  {Cornell}},\ }\bibfield  {title} {\enquote {\bibinfo {title} {Observation of
  bose-einstein condensation in a dilute atomic vapor},}\ }\href@noop {}
  {\bibfield  {journal} {\bibinfo  {journal} {science}\ }\textbf {\bibinfo
  {volume} {269}},\ \bibinfo {pages} {198--201} (\bibinfo {year}
  {1995})}\BibitemShut {NoStop}%
\bibitem [{\citenamefont {Bradley}\ \emph {et~al.}(1995)\citenamefont
  {Bradley}, \citenamefont {Sackett}, \citenamefont {Tollett},\ and\
  \citenamefont {Hulet}}]{bradley1995evidence}%
  \BibitemOpen
  \bibfield  {author} {\bibinfo {author} {\bibfnamefont {Cl~C}\ \bibnamefont
  {Bradley}}, \bibinfo {author} {\bibfnamefont {CA}~\bibnamefont {Sackett}},
  \bibinfo {author} {\bibfnamefont {JJ}~\bibnamefont {Tollett}}, \ and\
  \bibinfo {author} {\bibfnamefont {Randall~G}\ \bibnamefont {Hulet}},\
  }\bibfield  {title} {\enquote {\bibinfo {title} {Evidence of bose-einstein
  condensation in an atomic gas with attractive interactions},}\ }\href@noop {}
  {\bibfield  {journal} {\bibinfo  {journal} {Physical review letters}\
  }\textbf {\bibinfo {volume} {75}},\ \bibinfo {pages} {1687} (\bibinfo {year}
  {1995})}\BibitemShut {NoStop}%
\bibitem [{\citenamefont {Davis}\ \emph {et~al.}(1995)\citenamefont {Davis},
  \citenamefont {Mewes}, \citenamefont {Andrews}, \citenamefont {van Druten},
  \citenamefont {Durfee}, \citenamefont {Kurn},\ and\ \citenamefont
  {Ketterle}}]{davis1995bose}%
  \BibitemOpen
  \bibfield  {author} {\bibinfo {author} {\bibfnamefont {Kendall~B}\
  \bibnamefont {Davis}}, \bibinfo {author} {\bibfnamefont {M-O}\ \bibnamefont
  {Mewes}}, \bibinfo {author} {\bibfnamefont {Michael~R}\ \bibnamefont
  {Andrews}}, \bibinfo {author} {\bibfnamefont {Nicolaas~J}\ \bibnamefont {van
  Druten}}, \bibinfo {author} {\bibfnamefont {Dallin~S}\ \bibnamefont
  {Durfee}}, \bibinfo {author} {\bibfnamefont {DM}~\bibnamefont {Kurn}}, \ and\
  \bibinfo {author} {\bibfnamefont {Wolfgang}\ \bibnamefont {Ketterle}},\
  }\bibfield  {title} {\enquote {\bibinfo {title} {Bose-einstein condensation
  in a gas of sodium atoms},}\ }\href@noop {} {\bibfield  {journal} {\bibinfo
  {journal} {Physical review letters}\ }\textbf {\bibinfo {volume} {75}},\
  \bibinfo {pages} {3969} (\bibinfo {year} {1995})}\BibitemShut {NoStop}%
\bibitem [{\citenamefont {DeMarco}\ and\ \citenamefont
  {Jin}(1999)}]{demarco1999onset}%
  \BibitemOpen
  \bibfield  {author} {\bibinfo {author} {\bibfnamefont {Brian}\ \bibnamefont
  {DeMarco}}\ and\ \bibinfo {author} {\bibfnamefont {Deborah~S}\ \bibnamefont
  {Jin}},\ }\bibfield  {title} {\enquote {\bibinfo {title} {Onset of fermi
  degeneracy in a trapped atomic gas},}\ }\href@noop {} {\bibfield  {journal}
  {\bibinfo  {journal} {science}\ }\textbf {\bibinfo {volume} {285}},\ \bibinfo
  {pages} {1703--1706} (\bibinfo {year} {1999})}\BibitemShut {NoStop}%
\bibitem [{\citenamefont {Schreck}\ \emph {et~al.}(2001)\citenamefont
  {Schreck}, \citenamefont {Khaykovich}, \citenamefont {Corwin}, \citenamefont
  {Ferrari}, \citenamefont {Bourdel}, \citenamefont {Cubizolles},\ and\
  \citenamefont {Salomon}}]{schreck2001quasipure}%
  \BibitemOpen
  \bibfield  {author} {\bibinfo {author} {\bibfnamefont {F}~\bibnamefont
  {Schreck}}, \bibinfo {author} {\bibfnamefont {Lev}\ \bibnamefont
  {Khaykovich}}, \bibinfo {author} {\bibfnamefont {KL}~\bibnamefont {Corwin}},
  \bibinfo {author} {\bibfnamefont {G}~\bibnamefont {Ferrari}}, \bibinfo
  {author} {\bibfnamefont {Thomas}\ \bibnamefont {Bourdel}}, \bibinfo {author}
  {\bibfnamefont {Julien}\ \bibnamefont {Cubizolles}}, \ and\ \bibinfo {author}
  {\bibfnamefont {Christophe}\ \bibnamefont {Salomon}},\ }\bibfield  {title}
  {\enquote {\bibinfo {title} {Quasipure bose-einstein condensate immersed in a
  fermi sea},}\ }\href@noop {} {\bibfield  {journal} {\bibinfo  {journal}
  {Physical Review Letters}\ }\textbf {\bibinfo {volume} {87}},\ \bibinfo
  {pages} {080403} (\bibinfo {year} {2001})}\BibitemShut {NoStop}%
\bibitem [{\citenamefont {Truscott}\ \emph {et~al.}(2001)\citenamefont
  {Truscott}, \citenamefont {Strecker}, \citenamefont {McAlexander},
  \citenamefont {Partridge},\ and\ \citenamefont
  {Hulet}}]{truscott2001observation}%
  \BibitemOpen
  \bibfield  {author} {\bibinfo {author} {\bibfnamefont {Andrew~G}\
  \bibnamefont {Truscott}}, \bibinfo {author} {\bibfnamefont {Kevin~E}\
  \bibnamefont {Strecker}}, \bibinfo {author} {\bibfnamefont {William~I}\
  \bibnamefont {McAlexander}}, \bibinfo {author} {\bibfnamefont {Guthrie~B}\
  \bibnamefont {Partridge}}, \ and\ \bibinfo {author} {\bibfnamefont
  {Randall~G}\ \bibnamefont {Hulet}},\ }\bibfield  {title} {\enquote {\bibinfo
  {title} {Observation of fermi pressure in a gas of trapped atoms},}\
  }\href@noop {} {\bibfield  {journal} {\bibinfo  {journal} {Science}\ }\textbf
  {\bibinfo {volume} {291}},\ \bibinfo {pages} {2570--2572} (\bibinfo {year}
  {2001})}\BibitemShut {NoStop}%
\bibitem [{\citenamefont {O'hara}\ \emph {et~al.}(2002)\citenamefont {O'hara},
  \citenamefont {Hemmer}, \citenamefont {Gehm}, \citenamefont {Granade},\ and\
  \citenamefont {Thomas}}]{o2002observation}%
  \BibitemOpen
  \bibfield  {author} {\bibinfo {author} {\bibfnamefont {KM}~\bibnamefont
  {O'hara}}, \bibinfo {author} {\bibfnamefont {SL}~\bibnamefont {Hemmer}},
  \bibinfo {author} {\bibfnamefont {ME}~\bibnamefont {Gehm}}, \bibinfo {author}
  {\bibfnamefont {SR}~\bibnamefont {Granade}}, \ and\ \bibinfo {author}
  {\bibfnamefont {JE}~\bibnamefont {Thomas}},\ }\bibfield  {title} {\enquote
  {\bibinfo {title} {Observation of a strongly interacting degenerate fermi gas
  of atoms},}\ }\href@noop {} {\bibfield  {journal} {\bibinfo  {journal}
  {Science}\ }\textbf {\bibinfo {volume} {298}},\ \bibinfo {pages} {2179--2182}
  (\bibinfo {year} {2002})}\BibitemShut {NoStop}%
\bibitem [{\citenamefont {Greiner}\ \emph {et~al.}(2003)\citenamefont
  {Greiner}, \citenamefont {Regal},\ and\ \citenamefont
  {Jin}}]{greiner2003emergence}%
  \BibitemOpen
  \bibfield  {author} {\bibinfo {author} {\bibfnamefont {Markus}\ \bibnamefont
  {Greiner}}, \bibinfo {author} {\bibfnamefont {Cindy~A}\ \bibnamefont
  {Regal}}, \ and\ \bibinfo {author} {\bibfnamefont {Deborah~S}\ \bibnamefont
  {Jin}},\ }\bibfield  {title} {\enquote {\bibinfo {title} {Emergence of a
  molecular bose--einstein condensate from a fermi gas},}\ }\href@noop {}
  {\bibfield  {journal} {\bibinfo  {journal} {Nature}\ }\textbf {\bibinfo
  {volume} {426}},\ \bibinfo {pages} {537--540} (\bibinfo {year}
  {2003})}\BibitemShut {NoStop}%
\bibitem [{\citenamefont {Jochim}\ \emph {et~al.}(2003)\citenamefont {Jochim},
  \citenamefont {Bartenstein}, \citenamefont {Altmeyer}, \citenamefont {Hendl},
  \citenamefont {Riedl}, \citenamefont {Chin}, \citenamefont {Denschlag},\ and\
  \citenamefont {Grimm}}]{jochim2003bose}%
  \BibitemOpen
  \bibfield  {author} {\bibinfo {author} {\bibfnamefont {Selim}\ \bibnamefont
  {Jochim}}, \bibinfo {author} {\bibfnamefont {Markus}\ \bibnamefont
  {Bartenstein}}, \bibinfo {author} {\bibfnamefont {Alexander}\ \bibnamefont
  {Altmeyer}}, \bibinfo {author} {\bibfnamefont {Gerhard}\ \bibnamefont
  {Hendl}}, \bibinfo {author} {\bibfnamefont {Stefan}\ \bibnamefont {Riedl}},
  \bibinfo {author} {\bibfnamefont {Cheng}\ \bibnamefont {Chin}}, \bibinfo
  {author} {\bibfnamefont {J~Hecker}\ \bibnamefont {Denschlag}}, \ and\
  \bibinfo {author} {\bibfnamefont {Rudolf}\ \bibnamefont {Grimm}},\ }\bibfield
   {title} {\enquote {\bibinfo {title} {Bose-einstein condensation of
  molecules},}\ }\href@noop {} {\bibfield  {journal} {\bibinfo  {journal}
  {Science}\ }\textbf {\bibinfo {volume} {302}},\ \bibinfo {pages} {2101--2103}
  (\bibinfo {year} {2003})}\BibitemShut {NoStop}%
\bibitem [{\citenamefont {Zwierlein}\ \emph {et~al.}(2003)\citenamefont
  {Zwierlein}, \citenamefont {Hadzibabic}, \citenamefont {Gupta},\ and\
  \citenamefont {Ketterle}}]{zwierlein2003spectroscopic}%
  \BibitemOpen
  \bibfield  {author} {\bibinfo {author} {\bibfnamefont {Martin~W}\
  \bibnamefont {Zwierlein}}, \bibinfo {author} {\bibfnamefont {Zoran}\
  \bibnamefont {Hadzibabic}}, \bibinfo {author} {\bibfnamefont {Subhadeep}\
  \bibnamefont {Gupta}}, \ and\ \bibinfo {author} {\bibfnamefont {Wolfgang}\
  \bibnamefont {Ketterle}},\ }\bibfield  {title} {\enquote {\bibinfo {title}
  {Spectroscopic insensitivity to cold collisions in a two-state mixture of
  fermions},}\ }\href@noop {} {\bibfield  {journal} {\bibinfo  {journal}
  {Physical review letters}\ }\textbf {\bibinfo {volume} {91}},\ \bibinfo
  {pages} {250404} (\bibinfo {year} {2003})}\BibitemShut {NoStop}%
\bibitem [{\citenamefont {Greiner}\ \emph {et~al.}(2002)\citenamefont
  {Greiner}, \citenamefont {Mandel}, \citenamefont {Esslinger}, \citenamefont
  {H{\"a}nsch},\ and\ \citenamefont {Bloch}}]{grein}%
  \BibitemOpen
  \bibfield  {author} {\bibinfo {author} {\bibfnamefont {Markus}\ \bibnamefont
  {Greiner}}, \bibinfo {author} {\bibfnamefont {Olaf}\ \bibnamefont {Mandel}},
  \bibinfo {author} {\bibfnamefont {Tilman}\ \bibnamefont {Esslinger}},
  \bibinfo {author} {\bibfnamefont {Theodor~W}\ \bibnamefont {H{\"a}nsch}}, \
  and\ \bibinfo {author} {\bibfnamefont {Immanuel}\ \bibnamefont {Bloch}},\
  }\bibfield  {title} {\enquote {\bibinfo {title} {Quantum phase transition
  from a superfluid to a mott insulator in a gas of ultracold atoms},}\
  }\href@noop {} {\bibfield  {journal} {\bibinfo  {journal} {nature}\ }\textbf
  {\bibinfo {volume} {415}},\ \bibinfo {pages} {39--44} (\bibinfo {year}
  {2002})}\BibitemShut {NoStop}%
\bibitem [{\citenamefont {Lewenstein}\ \emph {et~al.}(2007)\citenamefont
  {Lewenstein}, \citenamefont {Sanpera}, \citenamefont {Ahufinger},
  \citenamefont {Damski}, \citenamefont {Sen},\ and\ \citenamefont
  {Sen}}]{lewenstein2007ultracold}%
  \BibitemOpen
  \bibfield  {author} {\bibinfo {author} {\bibfnamefont {Maciej}\ \bibnamefont
  {Lewenstein}}, \bibinfo {author} {\bibfnamefont {Anna}\ \bibnamefont
  {Sanpera}}, \bibinfo {author} {\bibfnamefont {Veronica}\ \bibnamefont
  {Ahufinger}}, \bibinfo {author} {\bibfnamefont {Bogdan}\ \bibnamefont
  {Damski}}, \bibinfo {author} {\bibfnamefont {Aditi}\ \bibnamefont {Sen}}, \
  and\ \bibinfo {author} {\bibfnamefont {Ujjwal}\ \bibnamefont {Sen}},\
  }\bibfield  {title} {\enquote {\bibinfo {title} {Ultracold atomic gases in
  optical lattices: mimicking condensed matter physics and beyond},}\
  }\href@noop {} {\bibfield  {journal} {\bibinfo  {journal} {Advances in
  Physics}\ }\textbf {\bibinfo {volume} {56}},\ \bibinfo {pages} {243--379}
  (\bibinfo {year} {2007})}\BibitemShut {NoStop}%
\bibitem [{\citenamefont {Lewenstein}\ \emph {et~al.}(2012)\citenamefont
  {Lewenstein}, \citenamefont {Sanpera},\ and\ \citenamefont
  {Ahufinger}}]{lewenstein2012ultracold}%
  \BibitemOpen
  \bibfield  {author} {\bibinfo {author} {\bibfnamefont {Maciej}\ \bibnamefont
  {Lewenstein}}, \bibinfo {author} {\bibfnamefont {Anna}\ \bibnamefont
  {Sanpera}}, \ and\ \bibinfo {author} {\bibfnamefont {Veronica}\ \bibnamefont
  {Ahufinger}},\ }\href@noop {} {\emph {\bibinfo {title} {Ultracold Atoms in
  Optical Lattices: Simulating quantum many-body systems}}}\ (\bibinfo
  {publisher} {Oxford University Press},\ \bibinfo {year} {2012})\BibitemShut
  {NoStop}%
\bibitem [{\citenamefont {Gross}\ and\ \citenamefont
  {Bloch}(2017)}]{gross2017quantum}%
  \BibitemOpen
  \bibfield  {author} {\bibinfo {author} {\bibfnamefont {Christian}\
  \bibnamefont {Gross}}\ and\ \bibinfo {author} {\bibfnamefont {Immanuel}\
  \bibnamefont {Bloch}},\ }\bibfield  {title} {\enquote {\bibinfo {title}
  {Quantum simulations with ultracold atoms in optical lattices},}\ }\href@noop
  {} {\bibfield  {journal} {\bibinfo  {journal} {Science}\ }\textbf {\bibinfo
  {volume} {357}},\ \bibinfo {pages} {995--1001} (\bibinfo {year}
  {2017})}\BibitemShut {NoStop}%
\bibitem [{\citenamefont {Martinez}\ \emph {et~al.}(2016)\citenamefont
  {Martinez}, \citenamefont {Muschik}, \citenamefont {Schindler}, \citenamefont
  {Nigg}, \citenamefont {Erhard}, \citenamefont {Heyl}, \citenamefont {Hauke},
  \citenamefont {Dalmonte}, \citenamefont {Monz}, \citenamefont {Zoller} \emph
  {et~al.}}]{martinez2016real}%
  \BibitemOpen
  \bibfield  {author} {\bibinfo {author} {\bibfnamefont {Esteban~A}\
  \bibnamefont {Martinez}}, \bibinfo {author} {\bibfnamefont {Christine~A}\
  \bibnamefont {Muschik}}, \bibinfo {author} {\bibfnamefont {Philipp}\
  \bibnamefont {Schindler}}, \bibinfo {author} {\bibfnamefont {Daniel}\
  \bibnamefont {Nigg}}, \bibinfo {author} {\bibfnamefont {Alexander}\
  \bibnamefont {Erhard}}, \bibinfo {author} {\bibfnamefont {Markus}\
  \bibnamefont {Heyl}}, \bibinfo {author} {\bibfnamefont {Philipp}\
  \bibnamefont {Hauke}}, \bibinfo {author} {\bibfnamefont {Marcello}\
  \bibnamefont {Dalmonte}}, \bibinfo {author} {\bibfnamefont {Thomas}\
  \bibnamefont {Monz}}, \bibinfo {author} {\bibfnamefont {Peter}\ \bibnamefont
  {Zoller}},  \emph {et~al.},\ }\bibfield  {title} {\enquote {\bibinfo {title}
  {Real-time dynamics of lattice gauge theories with a few-qubit quantum
  computer},}\ }\href@noop {} {\bibfield  {journal} {\bibinfo  {journal}
  {Nature}\ }\textbf {\bibinfo {volume} {534}},\ \bibinfo {pages} {516--519}
  (\bibinfo {year} {2016})}\BibitemShut {NoStop}%
\bibitem [{\citenamefont {Kogut}\ and\ \citenamefont
  {Susskind}(1975)}]{Kogut:1974ag}%
  \BibitemOpen
  \bibfield  {author} {\bibinfo {author} {\bibfnamefont {John~B.}\ \bibnamefont
  {Kogut}}\ and\ \bibinfo {author} {\bibfnamefont {Leonard}\ \bibnamefont
  {Susskind}},\ }\bibfield  {title} {\enquote {\bibinfo {title} {{Hamiltonian
  Formulation of Wilson's Lattice Gauge Theories}},}\ }\href {\doibase
  10.1103/PhysRevD.11.395} {\bibfield  {journal} {\bibinfo  {journal} {Phys.
  Rev. D}\ }\textbf {\bibinfo {volume} {11}},\ \bibinfo {pages} {395--408}
  (\bibinfo {year} {1975})}\BibitemShut {NoStop}%
\bibitem [{\citenamefont {Davoudi}\ \emph
  {et~al.}(2020{\natexlab{b}})\citenamefont {Davoudi}, \citenamefont
  {Raychowdhury},\ and\ \citenamefont {Shaw}}]{new}%
  \BibitemOpen
  \bibfield  {author} {\bibinfo {author} {\bibfnamefont {Zohreh}\ \bibnamefont
  {Davoudi}}, \bibinfo {author} {\bibfnamefont {Indrakshi}\ \bibnamefont
  {Raychowdhury}}, \ and\ \bibinfo {author} {\bibfnamefont {Andrew}\
  \bibnamefont {Shaw}},\ }\bibfield  {title} {\enquote {\bibinfo {title}
  {{Search for Efficient Formulations for Hamiltonian Simulation of non-Abelian
  Lattice Gauge Theories}},}\ }\href@noop {} {\  (\bibinfo {year}
  {2020}{\natexlab{b}})},\ \Eprint {http://arxiv.org/abs/2009.11802}
  {arXiv:2009.11802 [hep-lat]} \BibitemShut {NoStop}%
\bibitem [{\citenamefont {Chandrasekharan}\ and\ \citenamefont
  {Wiese}(1997)}]{chandrasekharan1997quantum}%
  \BibitemOpen
  \bibfield  {author} {\bibinfo {author} {\bibfnamefont {Shailesh}\
  \bibnamefont {Chandrasekharan}}\ and\ \bibinfo {author} {\bibfnamefont {U-J}\
  \bibnamefont {Wiese}},\ }\bibfield  {title} {\enquote {\bibinfo {title}
  {Quantum link models: A discrete approach to gauge theories},}\ }\href@noop
  {} {\bibfield  {journal} {\bibinfo  {journal} {Nuclear Physics B}\ }\textbf
  {\bibinfo {volume} {492}},\ \bibinfo {pages} {455--471} (\bibinfo {year}
  {1997})}\BibitemShut {NoStop}%
\bibitem [{\citenamefont {Brower}\ \emph {et~al.}(1999)\citenamefont {Brower},
  \citenamefont {Chandrasekharan},\ and\ \citenamefont
  {Wiese}}]{Brower:1997ha}%
  \BibitemOpen
  \bibfield  {author} {\bibinfo {author} {\bibfnamefont {R.}~\bibnamefont
  {Brower}}, \bibinfo {author} {\bibfnamefont {S.}~\bibnamefont
  {Chandrasekharan}}, \ and\ \bibinfo {author} {\bibfnamefont {U.J.}\
  \bibnamefont {Wiese}},\ }\bibfield  {title} {\enquote {\bibinfo {title} {{QCD
  as a quantum link model}},}\ }\href {\doibase 10.1103/PhysRevD.60.094502}
  {\bibfield  {journal} {\bibinfo  {journal} {Phys. Rev. D}\ }\textbf {\bibinfo
  {volume} {60}},\ \bibinfo {pages} {094502} (\bibinfo {year} {1999})},\
  \Eprint {http://arxiv.org/abs/hep-th/9704106} {arXiv:hep-th/9704106}
  \BibitemShut {NoStop}%
\bibitem [{\citenamefont {Zohar}\ and\ \citenamefont
  {Cirac}(2019)}]{Zohar:2019ygc}%
  \BibitemOpen
  \bibfield  {author} {\bibinfo {author} {\bibfnamefont {Erez}\ \bibnamefont
  {Zohar}}\ and\ \bibinfo {author} {\bibfnamefont {J.~Ignacio}\ \bibnamefont
  {Cirac}},\ }\bibfield  {title} {\enquote {\bibinfo {title} {{Removing
  Staggered Fermionic Matter in $U(N)$ and $SU(N)$ Lattice Gauge Theories}},}\
  }\href {\doibase 10.1103/PhysRevD.99.114511} {\bibfield  {journal} {\bibinfo
  {journal} {Phys. Rev. D}\ }\textbf {\bibinfo {volume} {99}},\ \bibinfo
  {pages} {114511} (\bibinfo {year} {2019})},\ \Eprint
  {http://arxiv.org/abs/1905.00652} {arXiv:1905.00652 [quant-ph]} \BibitemShut
  {NoStop}%
\bibitem [{\citenamefont {Zohar}\ and\ \citenamefont
  {Cirac}(2018)}]{Zohar:2018cwb}%
  \BibitemOpen
  \bibfield  {author} {\bibinfo {author} {\bibfnamefont {Erez}\ \bibnamefont
  {Zohar}}\ and\ \bibinfo {author} {\bibfnamefont {J.~Ignacio}\ \bibnamefont
  {Cirac}},\ }\bibfield  {title} {\enquote {\bibinfo {title} {{Eliminating
  fermionic matter fields in lattice gauge theories}},}\ }\href {\doibase
  10.1103/PhysRevB.98.075119} {\bibfield  {journal} {\bibinfo  {journal} {Phys.
  Rev. B}\ }\textbf {\bibinfo {volume} {98}},\ \bibinfo {pages} {075119}
  (\bibinfo {year} {2018})},\ \Eprint {http://arxiv.org/abs/1805.05347}
  {arXiv:1805.05347 [quant-ph]} \BibitemShut {NoStop}%
\bibitem [{\citenamefont {Zohar}\ and\ \citenamefont
  {Burrello}(2015)}]{Zohar:2014qma}%
  \BibitemOpen
  \bibfield  {author} {\bibinfo {author} {\bibfnamefont {Erez}\ \bibnamefont
  {Zohar}}\ and\ \bibinfo {author} {\bibfnamefont {Michele}\ \bibnamefont
  {Burrello}},\ }\bibfield  {title} {\enquote {\bibinfo {title} {{Formulation
  of lattice gauge theories for quantum simulations}},}\ }\href {\doibase
  10.1103/PhysRevD.91.054506} {\bibfield  {journal} {\bibinfo  {journal} {Phys.
  Rev. D}\ }\textbf {\bibinfo {volume} {91}},\ \bibinfo {pages} {054506}
  (\bibinfo {year} {2015})},\ \Eprint {http://arxiv.org/abs/1409.3085}
  {arXiv:1409.3085 [quant-ph]} \BibitemShut {NoStop}%
\bibitem [{\citenamefont {Ba{\~n}uls}\ \emph {et~al.}(2017)\citenamefont
  {Ba{\~n}uls}, \citenamefont {Cichy}, \citenamefont {Cirac}, \citenamefont
  {Jansen},\ and\ \citenamefont {K{\"u}hn}}]{banuls2017efficient}%
  \BibitemOpen
  \bibfield  {author} {\bibinfo {author} {\bibfnamefont {Mari~Carmen}\
  \bibnamefont {Ba{\~n}uls}}, \bibinfo {author} {\bibfnamefont {Krzysztof}\
  \bibnamefont {Cichy}}, \bibinfo {author} {\bibfnamefont {J~Ignacio}\
  \bibnamefont {Cirac}}, \bibinfo {author} {\bibfnamefont {Karl}\ \bibnamefont
  {Jansen}}, \ and\ \bibinfo {author} {\bibfnamefont {Stefan}\ \bibnamefont
  {K{\"u}hn}},\ }\bibfield  {title} {\enquote {\bibinfo {title} {Efficient
  basis formulation for (1+ 1)-dimensional su (2) lattice gauge theory:
  Spectral calculations with matrix product states},}\ }\href@noop {}
  {\bibfield  {journal} {\bibinfo  {journal} {Physical Review X}\ }\textbf
  {\bibinfo {volume} {7}},\ \bibinfo {pages} {041046} (\bibinfo {year}
  {2017})}\BibitemShut {NoStop}%
\bibitem [{\citenamefont {Raychowdhury}\ and\ \citenamefont
  {Stryker}(2020{\natexlab{a}})}]{Raychowdhury:2019iki}%
  \BibitemOpen
  \bibfield  {author} {\bibinfo {author} {\bibfnamefont {Indrakshi}\
  \bibnamefont {Raychowdhury}}\ and\ \bibinfo {author} {\bibfnamefont
  {Jesse~R.}\ \bibnamefont {Stryker}},\ }\bibfield  {title} {\enquote {\bibinfo
  {title} {{Loop, String, and Hadron Dynamics in SU(2) Hamiltonian Lattice
  Gauge Theories}},}\ }\href {\doibase 10.1103/PhysRevD.101.114502} {\bibfield
  {journal} {\bibinfo  {journal} {Phys. Rev. D}\ }\textbf {\bibinfo {volume}
  {101}},\ \bibinfo {pages} {114502} (\bibinfo {year} {2020}{\natexlab{a}})},\
  \Eprint {http://arxiv.org/abs/1912.06133} {arXiv:1912.06133 [hep-lat]}
  \BibitemShut {NoStop}%
\bibitem [{\citenamefont {Sala}\ \emph {et~al.}(2018)\citenamefont {Sala},
  \citenamefont {Shi}, \citenamefont {K{\"u}hn}, \citenamefont {Banuls},
  \citenamefont {Demler},\ and\ \citenamefont {Cirac}}]{sala2018variational}%
  \BibitemOpen
  \bibfield  {author} {\bibinfo {author} {\bibfnamefont {Pablo}\ \bibnamefont
  {Sala}}, \bibinfo {author} {\bibfnamefont {Tao}\ \bibnamefont {Shi}},
  \bibinfo {author} {\bibfnamefont {Stefan}\ \bibnamefont {K{\"u}hn}}, \bibinfo
  {author} {\bibfnamefont {Mari~Carmen}\ \bibnamefont {Banuls}}, \bibinfo
  {author} {\bibfnamefont {Eugene}\ \bibnamefont {Demler}}, \ and\ \bibinfo
  {author} {\bibfnamefont {Juan~Ignacio}\ \bibnamefont {Cirac}},\ }\bibfield
  {title} {\enquote {\bibinfo {title} {Variational study of u (1) and su (2)
  lattice gauge theories with gaussian states in 1+ 1 dimensions},}\
  }\href@noop {} {\bibfield  {journal} {\bibinfo  {journal} {Physical Review
  D}\ }\textbf {\bibinfo {volume} {98}},\ \bibinfo {pages} {034505} (\bibinfo
  {year} {2018})}\BibitemShut {NoStop}%
\bibitem [{\citenamefont {Raychowdhury}\ and\ \citenamefont
  {Stryker}(2020{\natexlab{b}})}]{Raychowdhury:2018osk}%
  \BibitemOpen
  \bibfield  {author} {\bibinfo {author} {\bibfnamefont {Indrakshi}\
  \bibnamefont {Raychowdhury}}\ and\ \bibinfo {author} {\bibfnamefont
  {Jesse~R.}\ \bibnamefont {Stryker}},\ }\bibfield  {title} {\enquote {\bibinfo
  {title} {{Solving Gauss's Law on Digital Quantum Computers with
  Loop-String-Hadron Digitization}},}\ }\href {\doibase
  10.1103/PhysRevResearch.2.033039} {\bibfield  {journal} {\bibinfo  {journal}
  {Phys. Rev. Res.}\ }\textbf {\bibinfo {volume} {2}},\ \bibinfo {pages}
  {033039} (\bibinfo {year} {2020}{\natexlab{b}})},\ \Eprint
  {http://arxiv.org/abs/1812.07554} {arXiv:1812.07554 [hep-lat]} \BibitemShut
  {NoStop}%
\bibitem [{\citenamefont {Yang}\ \emph
  {et~al.}(2020{\natexlab{b}})\citenamefont {Yang}, \citenamefont {Sun},
  \citenamefont {Ott}, \citenamefont {Wang}, \citenamefont {Zache},
  \citenamefont {Halimeh}, \citenamefont {Yuan}, \citenamefont {Hauke},\ and\
  \citenamefont {Pan}}]{Yang:2020yer}%
  \BibitemOpen
  \bibfield  {author} {\bibinfo {author} {\bibfnamefont {Bing}\ \bibnamefont
  {Yang}}, \bibinfo {author} {\bibfnamefont {Hui}\ \bibnamefont {Sun}},
  \bibinfo {author} {\bibfnamefont {Robert}\ \bibnamefont {Ott}}, \bibinfo
  {author} {\bibfnamefont {Han-Yi}\ \bibnamefont {Wang}}, \bibinfo {author}
  {\bibfnamefont {Torsten~V.}\ \bibnamefont {Zache}}, \bibinfo {author}
  {\bibfnamefont {Jad~C.}\ \bibnamefont {Halimeh}}, \bibinfo {author}
  {\bibfnamefont {Zhen-Sheng}\ \bibnamefont {Yuan}}, \bibinfo {author}
  {\bibfnamefont {Philipp}\ \bibnamefont {Hauke}}, \ and\ \bibinfo {author}
  {\bibfnamefont {Jian-Wei}\ \bibnamefont {Pan}},\ }\bibfield  {title}
  {\enquote {\bibinfo {title} {{Observation of gauge invariance in a 71-site
  quantum simulator}},}\ }\href@noop {} {\  (\bibinfo {year}
  {2020}{\natexlab{b}})},\ \Eprint {http://arxiv.org/abs/2003.08945}
  {arXiv:2003.08945 [cond-mat.quant-gas]} \BibitemShut {NoStop}%
\bibitem [{\citenamefont {Mathur}(2005)}]{Mathur:2004kr}%
  \BibitemOpen
  \bibfield  {author} {\bibinfo {author} {\bibfnamefont {Manu}\ \bibnamefont
  {Mathur}},\ }\bibfield  {title} {\enquote {\bibinfo {title} {{Harmonic
  oscillator prepotentials in SU(2) lattice gauge theory}},}\ }\href {\doibase
  10.1088/0305-4470/38/46/008} {\bibfield  {journal} {\bibinfo  {journal} {J.
  Phys. A}\ }\textbf {\bibinfo {volume} {38}},\ \bibinfo {pages} {10015--10026}
  (\bibinfo {year} {2005})},\ \Eprint {http://arxiv.org/abs/hep-lat/0403029}
  {arXiv:hep-lat/0403029} \BibitemShut {NoStop}%
\bibitem [{\citenamefont {Mathur}(2007)}]{Mathur:2007nu}%
  \BibitemOpen
  \bibfield  {author} {\bibinfo {author} {\bibfnamefont {Manu}\ \bibnamefont
  {Mathur}},\ }\bibfield  {title} {\enquote {\bibinfo {title} {{Loop Approach
  to Lattice Gauge Theories}},}\ }\href {\doibase
  10.1016/j.nuclphysb.2007.04.031} {\bibfield  {journal} {\bibinfo  {journal}
  {Nucl. Phys. B}\ }\textbf {\bibinfo {volume} {779}},\ \bibinfo {pages}
  {32--62} (\bibinfo {year} {2007})},\ \Eprint
  {http://arxiv.org/abs/hep-lat/0702007} {arXiv:hep-lat/0702007} \BibitemShut
  {NoStop}%
\bibitem [{\citenamefont {Mathur}\ \emph {et~al.}(2010)\citenamefont {Mathur},
  \citenamefont {Raychowdhury},\ and\ \citenamefont
  {Anishetty}}]{Mathur:2010wc}%
  \BibitemOpen
  \bibfield  {author} {\bibinfo {author} {\bibfnamefont {Manu}\ \bibnamefont
  {Mathur}}, \bibinfo {author} {\bibfnamefont {Indrakshi}\ \bibnamefont
  {Raychowdhury}}, \ and\ \bibinfo {author} {\bibfnamefont {Ramesh}\
  \bibnamefont {Anishetty}},\ }\bibfield  {title} {\enquote {\bibinfo {title}
  {{SU(N) Irreducible Schwinger Bosons}},}\ }\href {\doibase 10.1063/1.3464267}
  {\bibfield  {journal} {\bibinfo  {journal} {J. Math. Phys.}\ }\textbf
  {\bibinfo {volume} {51}},\ \bibinfo {pages} {093504} (\bibinfo {year}
  {2010})},\ \Eprint {http://arxiv.org/abs/1003.5487} {arXiv:1003.5487
  [math-ph]} \BibitemShut {NoStop}%
\bibitem [{\citenamefont {Anishetty}\ \emph {et~al.}(2009)\citenamefont
  {Anishetty}, \citenamefont {Mathur},\ and\ \citenamefont
  {Raychowdhury}}]{Anishetty:2009ai}%
  \BibitemOpen
  \bibfield  {author} {\bibinfo {author} {\bibfnamefont {Ramesh}\ \bibnamefont
  {Anishetty}}, \bibinfo {author} {\bibfnamefont {Manu}\ \bibnamefont
  {Mathur}}, \ and\ \bibinfo {author} {\bibfnamefont {Indrakshi}\ \bibnamefont
  {Raychowdhury}},\ }\bibfield  {title} {\enquote {\bibinfo {title}
  {{Irreducible SU(3) Schhwinger Bosons}},}\ }\href {\doibase
  10.1063/1.3122666} {\bibfield  {journal} {\bibinfo  {journal} {J. Math.
  Phys.}\ }\textbf {\bibinfo {volume} {50}},\ \bibinfo {pages} {053503}
  (\bibinfo {year} {2009})},\ \Eprint {http://arxiv.org/abs/0901.0644}
  {arXiv:0901.0644 [math-ph]} \BibitemShut {NoStop}%
\bibitem [{\citenamefont {Anishetty}\ \emph {et~al.}(2010)\citenamefont
  {Anishetty}, \citenamefont {Mathur},\ and\ \citenamefont
  {Raychowdhury}}]{Anishetty:2009nh}%
  \BibitemOpen
  \bibfield  {author} {\bibinfo {author} {\bibfnamefont {Ramesh}\ \bibnamefont
  {Anishetty}}, \bibinfo {author} {\bibfnamefont {Manu}\ \bibnamefont
  {Mathur}}, \ and\ \bibinfo {author} {\bibfnamefont {Indrakshi}\ \bibnamefont
  {Raychowdhury}},\ }\bibfield  {title} {\enquote {\bibinfo {title}
  {{Prepotential formulation of SU(3) lattice gauge theory}},}\ }\href
  {\doibase 10.1088/1751-8113/43/3/035403} {\bibfield  {journal} {\bibinfo
  {journal} {J. Phys. A}\ }\textbf {\bibinfo {volume} {43}},\ \bibinfo {pages}
  {035403} (\bibinfo {year} {2010})},\ \Eprint {http://arxiv.org/abs/0909.2394}
  {arXiv:0909.2394 [hep-lat]} \BibitemShut {NoStop}%
\bibitem [{\citenamefont {Anishetty}\ and\ \citenamefont
  {Raychowdhury}(2014)}]{Anishetty:2014tta}%
  \BibitemOpen
  \bibfield  {author} {\bibinfo {author} {\bibfnamefont {Ramesh}\ \bibnamefont
  {Anishetty}}\ and\ \bibinfo {author} {\bibfnamefont {Indrakshi}\ \bibnamefont
  {Raychowdhury}},\ }\bibfield  {title} {\enquote {\bibinfo {title} {{SU(2)
  lattice gauge theory: Local dynamics on nonintersecting electric flux
  loops}},}\ }\href {\doibase 10.1103/PhysRevD.90.114503} {\bibfield  {journal}
  {\bibinfo  {journal} {Phys. Rev. D}\ }\textbf {\bibinfo {volume} {90}},\
  \bibinfo {pages} {114503} (\bibinfo {year} {2014})},\ \Eprint
  {http://arxiv.org/abs/1408.6331} {arXiv:1408.6331 [hep-lat]} \BibitemShut
  {NoStop}%
\bibitem [{\citenamefont {Raychowdhury}(2013)}]{Raychowdhury:2013rwa}%
  \BibitemOpen
  \bibfield  {author} {\bibinfo {author} {\bibfnamefont {Indrakshi}\
  \bibnamefont {Raychowdhury}},\ }\emph {\bibinfo {title} {{Prepotential
  Formulation of Lattice Gauge Theories}}},\ \href@noop {} {Ph.D. thesis},\
  \bibinfo  {school} {Calcutta U.} (\bibinfo {year} {2013})\BibitemShut
  {NoStop}%
\bibitem [{\citenamefont {Raychowdhury}\ and\ \citenamefont
  {Anishetty}(2014)}]{Raychowdhury:2014eta}%
  \BibitemOpen
  \bibfield  {author} {\bibinfo {author} {\bibfnamefont {Indrakshi}\
  \bibnamefont {Raychowdhury}}\ and\ \bibinfo {author} {\bibfnamefont {Ramesh}\
  \bibnamefont {Anishetty}},\ }\bibfield  {title} {\enquote {\bibinfo {title}
  {{Prepotential Formulation of Lattice Gauge Theory}},}\ }\href {\doibase
  10.22323/1.214.0313} {\bibfield  {journal} {\bibinfo  {journal} {PoS}\
  }\textbf {\bibinfo {volume} {LATTICE2014}},\ \bibinfo {pages} {313} (\bibinfo
  {year} {2014})},\ \Eprint {http://arxiv.org/abs/1411.3068} {arXiv:1411.3068
  [hep-lat]} \BibitemShut {NoStop}%
\bibitem [{\citenamefont {Raychowdhury}(2019)}]{Raychowdhury:2018tfj}%
  \BibitemOpen
  \bibfield  {author} {\bibinfo {author} {\bibfnamefont {Indrakshi}\
  \bibnamefont {Raychowdhury}},\ }\bibfield  {title} {\enquote {\bibinfo
  {title} {{Low energy spectrum of SU(2) lattice gauge theory}: {An alternate
  proposal via loop formulation}},}\ }\href {\doibase
  10.1140/epjc/s10052-019-6753-0} {\bibfield  {journal} {\bibinfo  {journal}
  {Eur. Phys. J. C}\ }\textbf {\bibinfo {volume} {79}},\ \bibinfo {pages} {235}
  (\bibinfo {year} {2019})},\ \Eprint {http://arxiv.org/abs/1804.01304}
  {arXiv:1804.01304 [hep-lat]} \BibitemShut {NoStop}%
\bibitem [{\citenamefont {Nagaosa}\ and\ \citenamefont
  {Takimoto}(1986)}]{nagaosa}%
  \BibitemOpen
  \bibfield  {author} {\bibinfo {author} {\bibfnamefont {Naoto}\ \bibnamefont
  {Nagaosa}}\ and\ \bibinfo {author} {\bibfnamefont {Jun-ichi}\ \bibnamefont
  {Takimoto}},\ }\bibfield  {title} {\enquote {\bibinfo {title} {Theory of
  neutral-ionic transition in organic crystals. i. monte carlo simulation of
  modified hubbard model},}\ }\href@noop {} {\bibfield  {journal} {\bibinfo
  {journal} {Journal of the Physical Society of Japan}\ }\textbf {\bibinfo
  {volume} {55}},\ \bibinfo {pages} {2735--2744} (\bibinfo {year}
  {1986})}\BibitemShut {NoStop}%
\bibitem [{\citenamefont {Egami}\ \emph {et~al.}(1993)\citenamefont {Egami},
  \citenamefont {Ishihara},\ and\ \citenamefont {Tachiki}}]{egami}%
  \BibitemOpen
  \bibfield  {author} {\bibinfo {author} {\bibfnamefont {T}~\bibnamefont
  {Egami}}, \bibinfo {author} {\bibfnamefont {S}~\bibnamefont {Ishihara}}, \
  and\ \bibinfo {author} {\bibfnamefont {M}~\bibnamefont {Tachiki}},\
  }\bibfield  {title} {\enquote {\bibinfo {title} {Lattice effect of strong
  electron correlation: Implication for ferroelectricity and
  superconductivity},}\ }\href@noop {} {\bibfield  {journal} {\bibinfo
  {journal} {Science}\ }\textbf {\bibinfo {volume} {261}},\ \bibinfo {pages}
  {1307--1310} (\bibinfo {year} {1993})}\BibitemShut {NoStop}%
\bibitem [{\citenamefont {Messer}\ \emph {et~al.}(2015)\citenamefont {Messer},
  \citenamefont {Desbuquois}, \citenamefont {Uehlinger}, \citenamefont {Jotzu},
  \citenamefont {Huber}, \citenamefont {Greif},\ and\ \citenamefont
  {Esslinger}}]{messer}%
  \BibitemOpen
  \bibfield  {author} {\bibinfo {author} {\bibfnamefont {Michael}\ \bibnamefont
  {Messer}}, \bibinfo {author} {\bibfnamefont {R{\'e}mi}\ \bibnamefont
  {Desbuquois}}, \bibinfo {author} {\bibfnamefont {Thomas}\ \bibnamefont
  {Uehlinger}}, \bibinfo {author} {\bibfnamefont {Gregor}\ \bibnamefont
  {Jotzu}}, \bibinfo {author} {\bibfnamefont {Sebastian}\ \bibnamefont
  {Huber}}, \bibinfo {author} {\bibfnamefont {Daniel}\ \bibnamefont {Greif}}, \
  and\ \bibinfo {author} {\bibfnamefont {Tilman}\ \bibnamefont {Esslinger}},\
  }\bibfield  {title} {\enquote {\bibinfo {title} {Exploring competing density
  order in the ionic hubbard model with ultracold fermions},}\ }\href@noop {}
  {\bibfield  {journal} {\bibinfo  {journal} {Physical review letters}\
  }\textbf {\bibinfo {volume} {115}},\ \bibinfo {pages} {115303} (\bibinfo
  {year} {2015})}\BibitemShut {NoStop}%
\bibitem [{\citenamefont {Fabrizio}\ \emph {et~al.}(1999)\citenamefont
  {Fabrizio}, \citenamefont {Gogolin},\ and\ \citenamefont
  {Nersesyan}}]{fabrizio1999band}%
  \BibitemOpen
  \bibfield  {author} {\bibinfo {author} {\bibfnamefont {Michele}\ \bibnamefont
  {Fabrizio}}, \bibinfo {author} {\bibfnamefont {Alexander~O}\ \bibnamefont
  {Gogolin}}, \ and\ \bibinfo {author} {\bibfnamefont {Alexander~A}\
  \bibnamefont {Nersesyan}},\ }\bibfield  {title} {\enquote {\bibinfo {title}
  {From band insulator to mott insulator in one dimension},}\ }\href@noop {}
  {\bibfield  {journal} {\bibinfo  {journal} {Physical review letters}\
  }\textbf {\bibinfo {volume} {83}},\ \bibinfo {pages} {2014} (\bibinfo {year}
  {1999})}\BibitemShut {NoStop}%
\bibitem [{\citenamefont {Kampf}\ \emph {et~al.}(2003)\citenamefont {Kampf},
  \citenamefont {Sekania}, \citenamefont {Japaridze},\ and\ \citenamefont
  {Brune}}]{kampf2003nature}%
  \BibitemOpen
  \bibfield  {author} {\bibinfo {author} {\bibfnamefont {Arno~P}\ \bibnamefont
  {Kampf}}, \bibinfo {author} {\bibfnamefont {Michael}\ \bibnamefont
  {Sekania}}, \bibinfo {author} {\bibfnamefont {George~I}\ \bibnamefont
  {Japaridze}}, \ and\ \bibinfo {author} {\bibfnamefont {Ph}~\bibnamefont
  {Brune}},\ }\bibfield  {title} {\enquote {\bibinfo {title} {Nature of the
  insulating phases in the half-filled ionic hubbard model},}\ }\href@noop {}
  {\bibfield  {journal} {\bibinfo  {journal} {Journal of Physics: Condensed
  Matter}\ }\textbf {\bibinfo {volume} {15}},\ \bibinfo {pages} {5895}
  (\bibinfo {year} {2003})}\BibitemShut {NoStop}%
\bibitem [{\citenamefont {Bag}\ \emph {et~al.}(2015)\citenamefont {Bag},
  \citenamefont {Garg},\ and\ \citenamefont {Krishnamurthy}}]{bag2015phase}%
  \BibitemOpen
  \bibfield  {author} {\bibinfo {author} {\bibfnamefont {Soumen}\ \bibnamefont
  {Bag}}, \bibinfo {author} {\bibfnamefont {Arti}\ \bibnamefont {Garg}}, \ and\
  \bibinfo {author} {\bibfnamefont {HR}~\bibnamefont {Krishnamurthy}},\
  }\bibfield  {title} {\enquote {\bibinfo {title} {Phase diagram of the
  half-filled ionic hubbard model},}\ }\href@noop {} {\bibfield  {journal}
  {\bibinfo  {journal} {Physical Review B}\ }\textbf {\bibinfo {volume} {91}},\
  \bibinfo {pages} {235108} (\bibinfo {year} {2015})}\BibitemShut {NoStop}%
\bibitem [{\citenamefont {Samanta}\ and\ \citenamefont
  {Sensarma}(2016)}]{samanta2016superconductivity}%
  \BibitemOpen
  \bibfield  {author} {\bibinfo {author} {\bibfnamefont {Abhisek}\ \bibnamefont
  {Samanta}}\ and\ \bibinfo {author} {\bibfnamefont {Rajdeep}\ \bibnamefont
  {Sensarma}},\ }\bibfield  {title} {\enquote {\bibinfo {title}
  {Superconductivity from doublon condensation in the ionic hubbard model},}\
  }\href@noop {} {\bibfield  {journal} {\bibinfo  {journal} {Physical Review
  B}\ }\textbf {\bibinfo {volume} {94}},\ \bibinfo {pages} {224517} (\bibinfo
  {year} {2016})}\BibitemShut {NoStop}%
\bibitem [{\citenamefont {Hamer}(1982)}]{Hamer:1981yq}%
  \BibitemOpen
  \bibfield  {author} {\bibinfo {author} {\bibfnamefont {C.J.}\ \bibnamefont
  {Hamer}},\ }\bibfield  {title} {\enquote {\bibinfo {title} {{SU(2)
  {Yang-Mills} Theory in (1+1)-dimensions: A Finite Lattice Approach}},}\
  }\href {\doibase 10.1016/0550-3213(82)90009-8} {\bibfield  {journal}
  {\bibinfo  {journal} {Nucl. Phys. B}\ }\textbf {\bibinfo {volume} {195}},\
  \bibinfo {pages} {503--521} (\bibinfo {year} {1982})}\BibitemShut {NoStop}%
\bibitem [{\citenamefont {Di~Liberto}\ \emph {et~al.}(2014)\citenamefont
  {Di~Liberto}, \citenamefont {Comparin}, \citenamefont {Kock}, \citenamefont
  {{\"O}lschl{\"a}ger}, \citenamefont {Hemmerich},\ and\ \citenamefont
  {Smith}}]{liberto}%
  \BibitemOpen
  \bibfield  {author} {\bibinfo {author} {\bibfnamefont {Marco}\ \bibnamefont
  {Di~Liberto}}, \bibinfo {author} {\bibfnamefont {Tommaso}\ \bibnamefont
  {Comparin}}, \bibinfo {author} {\bibfnamefont {Thorge}\ \bibnamefont {Kock}},
  \bibinfo {author} {\bibfnamefont {M}~\bibnamefont {{\"O}lschl{\"a}ger}},
  \bibinfo {author} {\bibfnamefont {Andreas}\ \bibnamefont {Hemmerich}}, \ and\
  \bibinfo {author} {\bibfnamefont {C~Morais}\ \bibnamefont {Smith}},\
  }\bibfield  {title} {\enquote {\bibinfo {title} {Controlling coherence via
  tuning of the population imbalance in a bipartite optical lattice},}\
  }\href@noop {} {\bibfield  {journal} {\bibinfo  {journal} {Nature
  communications}\ }\textbf {\bibinfo {volume} {5}},\ \bibinfo {pages} {1--6}
  (\bibinfo {year} {2014})}\BibitemShut {NoStop}%
\bibitem [{\citenamefont {Scherg}\ \emph {et~al.}(2018)\citenamefont {Scherg},
  \citenamefont {Kohlert}, \citenamefont {Herbrych}, \citenamefont {Stolpp},
  \citenamefont {Bordia}, \citenamefont {Schneider}, \citenamefont
  {Heidrich-Meisner}, \citenamefont {Bloch},\ and\ \citenamefont
  {Aidelsburger}}]{scherg}%
  \BibitemOpen
  \bibfield  {author} {\bibinfo {author} {\bibfnamefont {Sebastian}\
  \bibnamefont {Scherg}}, \bibinfo {author} {\bibfnamefont {Thomas}\
  \bibnamefont {Kohlert}}, \bibinfo {author} {\bibfnamefont {J}~\bibnamefont
  {Herbrych}}, \bibinfo {author} {\bibfnamefont {J}~\bibnamefont {Stolpp}},
  \bibinfo {author} {\bibfnamefont {Pranjal}\ \bibnamefont {Bordia}}, \bibinfo
  {author} {\bibfnamefont {Ulrich}\ \bibnamefont {Schneider}}, \bibinfo
  {author} {\bibfnamefont {Fabian}\ \bibnamefont {Heidrich-Meisner}}, \bibinfo
  {author} {\bibfnamefont {Immanuel}\ \bibnamefont {Bloch}}, \ and\ \bibinfo
  {author} {\bibfnamefont {Monika}\ \bibnamefont {Aidelsburger}},\ }\bibfield
  {title} {\enquote {\bibinfo {title} {Nonequilibrium mass transport in the 1d
  fermi-hubbard model},}\ }\href@noop {} {\bibfield  {journal} {\bibinfo
  {journal} {Physical Review Letters}\ }\textbf {\bibinfo {volume} {121}},\
  \bibinfo {pages} {130402} (\bibinfo {year} {2018})}\BibitemShut {NoStop}%
\bibitem [{\citenamefont {Ronzheimer}\ \emph {et~al.}(2013)\citenamefont
  {Ronzheimer}, \citenamefont {Schreiber}, \citenamefont {Braun}, \citenamefont
  {Hodgman}, \citenamefont {Langer}, \citenamefont {McCulloch}, \citenamefont
  {Heidrich-Meisner}, \citenamefont {Bloch},\ and\ \citenamefont
  {Schneider}}]{ronz}%
  \BibitemOpen
  \bibfield  {author} {\bibinfo {author} {\bibfnamefont {J.~P.}\ \bibnamefont
  {Ronzheimer}}, \bibinfo {author} {\bibfnamefont {M.}~\bibnamefont
  {Schreiber}}, \bibinfo {author} {\bibfnamefont {S.}~\bibnamefont {Braun}},
  \bibinfo {author} {\bibfnamefont {S.~S.}\ \bibnamefont {Hodgman}}, \bibinfo
  {author} {\bibfnamefont {S.}~\bibnamefont {Langer}}, \bibinfo {author}
  {\bibfnamefont {I.~P.}\ \bibnamefont {McCulloch}}, \bibinfo {author}
  {\bibfnamefont {F.}~\bibnamefont {Heidrich-Meisner}}, \bibinfo {author}
  {\bibfnamefont {I.}~\bibnamefont {Bloch}}, \ and\ \bibinfo {author}
  {\bibfnamefont {U.}~\bibnamefont {Schneider}},\ }\bibfield  {title} {\enquote
  {\bibinfo {title} {Expansion dynamics of interacting bosons in homogeneous
  lattices in one and two dimensions},}\ }\href {\doibase
  10.1103/PhysRevLett.110.205301} {\bibfield  {journal} {\bibinfo  {journal}
  {Phys. Rev. Lett.}\ }\textbf {\bibinfo {volume} {110}},\ \bibinfo {pages}
  {205301} (\bibinfo {year} {2013})}\BibitemShut {NoStop}%
\bibitem [{\citenamefont {Sponselee}\ \emph {et~al.}(2018)\citenamefont
  {Sponselee}, \citenamefont {Freystatzky}, \citenamefont {Abeln},
  \citenamefont {Diem}, \citenamefont {Hundt}, \citenamefont {Kochanke},
  \citenamefont {Ponath}, \citenamefont {Santra}, \citenamefont {Mathey},
  \citenamefont {Sengstock} \emph {et~al.}}]{spon}%
  \BibitemOpen
  \bibfield  {author} {\bibinfo {author} {\bibfnamefont {Koen}\ \bibnamefont
  {Sponselee}}, \bibinfo {author} {\bibfnamefont {Lukas}\ \bibnamefont
  {Freystatzky}}, \bibinfo {author} {\bibfnamefont {Benjamin}\ \bibnamefont
  {Abeln}}, \bibinfo {author} {\bibfnamefont {Marcel}\ \bibnamefont {Diem}},
  \bibinfo {author} {\bibfnamefont {Bastian}\ \bibnamefont {Hundt}}, \bibinfo
  {author} {\bibfnamefont {Andr{\'e}}\ \bibnamefont {Kochanke}}, \bibinfo
  {author} {\bibfnamefont {Thomas}\ \bibnamefont {Ponath}}, \bibinfo {author}
  {\bibfnamefont {Bodhaditya}\ \bibnamefont {Santra}}, \bibinfo {author}
  {\bibfnamefont {Ludwig}\ \bibnamefont {Mathey}}, \bibinfo {author}
  {\bibfnamefont {Klaus}\ \bibnamefont {Sengstock}},  \emph {et~al.},\
  }\bibfield  {title} {\enquote {\bibinfo {title} {Dynamics of ultracold
  quantum gases in the dissipative fermi--hubbard model},}\ }\href@noop {}
  {\bibfield  {journal} {\bibinfo  {journal} {Quantum Science and Technology}\
  }\textbf {\bibinfo {volume} {4}},\ \bibinfo {pages} {014002} (\bibinfo {year}
  {2018})}\BibitemShut {NoStop}%
\bibitem [{\citenamefont {Bloch}\ \emph {et~al.}(2008)\citenamefont {Bloch},
  \citenamefont {Dalibard},\ and\ \citenamefont {Zwerger}}]{bloch}%
  \BibitemOpen
  \bibfield  {author} {\bibinfo {author} {\bibfnamefont {Immanuel}\
  \bibnamefont {Bloch}}, \bibinfo {author} {\bibfnamefont {Jean}\ \bibnamefont
  {Dalibard}}, \ and\ \bibinfo {author} {\bibfnamefont {Wilhelm}\ \bibnamefont
  {Zwerger}},\ }\bibfield  {title} {\enquote {\bibinfo {title} {Many-body
  physics with ultracold gases},}\ }\href@noop {} {\bibfield  {journal}
  {\bibinfo  {journal} {Reviews of modern physics}\ }\textbf {\bibinfo {volume}
  {80}},\ \bibinfo {pages} {885} (\bibinfo {year} {2008})}\BibitemShut
  {NoStop}%
\bibitem [{\citenamefont {Schreiber}\ \emph {et~al.}(2015)\citenamefont
  {Schreiber}, \citenamefont {Hodgman}, \citenamefont {Bordia}, \citenamefont
  {L{\"u}schen}, \citenamefont {Fischer}, \citenamefont {Vosk}, \citenamefont
  {Altman}, \citenamefont {Schneider},\ and\ \citenamefont
  {Bloch}}]{schreiber}%
  \BibitemOpen
  \bibfield  {author} {\bibinfo {author} {\bibfnamefont {Michael}\ \bibnamefont
  {Schreiber}}, \bibinfo {author} {\bibfnamefont {Sean~S}\ \bibnamefont
  {Hodgman}}, \bibinfo {author} {\bibfnamefont {Pranjal}\ \bibnamefont
  {Bordia}}, \bibinfo {author} {\bibfnamefont {Henrik~P}\ \bibnamefont
  {L{\"u}schen}}, \bibinfo {author} {\bibfnamefont {Mark~H}\ \bibnamefont
  {Fischer}}, \bibinfo {author} {\bibfnamefont {Ronen}\ \bibnamefont {Vosk}},
  \bibinfo {author} {\bibfnamefont {Ehud}\ \bibnamefont {Altman}}, \bibinfo
  {author} {\bibfnamefont {Ulrich}\ \bibnamefont {Schneider}}, \ and\ \bibinfo
  {author} {\bibfnamefont {Immanuel}\ \bibnamefont {Bloch}},\ }\bibfield
  {title} {\enquote {\bibinfo {title} {Observation of many-body localization of
  interacting fermions in a quasirandom optical lattice},}\ }\href@noop {}
  {\bibfield  {journal} {\bibinfo  {journal} {Science}\ }\textbf {\bibinfo
  {volume} {349}},\ \bibinfo {pages} {842--845} (\bibinfo {year}
  {2015})}\BibitemShut {NoStop}%
\bibitem [{\citenamefont {Aidelsburger}\ \emph {et~al.}(2013)\citenamefont
  {Aidelsburger}, \citenamefont {Atala}, \citenamefont {Lohse}, \citenamefont
  {Barreiro}, \citenamefont {Paredes},\ and\ \citenamefont {Bloch}}]{aidel}%
  \BibitemOpen
  \bibfield  {author} {\bibinfo {author} {\bibfnamefont {Monika}\ \bibnamefont
  {Aidelsburger}}, \bibinfo {author} {\bibfnamefont {Marcos}\ \bibnamefont
  {Atala}}, \bibinfo {author} {\bibfnamefont {Michael}\ \bibnamefont {Lohse}},
  \bibinfo {author} {\bibfnamefont {Julio~T}\ \bibnamefont {Barreiro}},
  \bibinfo {author} {\bibfnamefont {B}~\bibnamefont {Paredes}}, \ and\ \bibinfo
  {author} {\bibfnamefont {Immanuel}\ \bibnamefont {Bloch}},\ }\bibfield
  {title} {\enquote {\bibinfo {title} {Realization of the hofstadter
  hamiltonian with ultracold atoms in optical lattices},}\ }\href@noop {}
  {\bibfield  {journal} {\bibinfo  {journal} {Physical review letters}\
  }\textbf {\bibinfo {volume} {111}},\ \bibinfo {pages} {185301} (\bibinfo
  {year} {2013})}\BibitemShut {NoStop}%
\bibitem [{\citenamefont {Imada}\ \emph {et~al.}(1998)\citenamefont {Imada},
  \citenamefont {Fujimori},\ and\ \citenamefont {Tokura}}]{imada1998metal}%
  \BibitemOpen
  \bibfield  {author} {\bibinfo {author} {\bibfnamefont {Masatoshi}\
  \bibnamefont {Imada}}, \bibinfo {author} {\bibfnamefont {Atsushi}\
  \bibnamefont {Fujimori}}, \ and\ \bibinfo {author} {\bibfnamefont
  {Yoshinori}\ \bibnamefont {Tokura}},\ }\bibfield  {title} {\enquote {\bibinfo
  {title} {Metal-insulator transitions},}\ }\href@noop {} {\bibfield  {journal}
  {\bibinfo  {journal} {Reviews of modern physics}\ }\textbf {\bibinfo {volume}
  {70}},\ \bibinfo {pages} {1039} (\bibinfo {year} {1998})}\BibitemShut
  {NoStop}%
\end{thebibliography}%
\appendix

\section{Approximate LSH Hamiltonian in the weak coupling limit}
\label{App:wcHamiltonian}
\noindent

In this appendix we derive the approximate Hamiltonian given in (\ref{wcHELSH},\ref{wcHMLSH},\ref{wcHILSH}) starting from the Hamiltonian given in (\ref{HELSH},\ref{HMLSH},\ref{HILSH}). 

\textbf{Electric Hamiltonian:} The electric part of the LSH Hamiltonian as given in (\ref{HELSH}) can be written as:
\bea
H_E^{(\mathrm{LSH})}=\frac{g^2a}{2}\sum_{j} h_E(j)
\eea
At each site $j$, depending upon the fermionic quantum numbers $n_i,n_o$, the local contribution to electric energy is given by,
\bea
\label{hE}
\begin{array}{c|c|c}
    n_i & n_o&h_E \\
    \hline && \\
     0&0 &\frac{n_l}{2}\left( \frac{n_l}{2}+1 \right)  \\  && \\
     0 &1 & \frac{n_l+1}{2}\left( \frac{n_l+1}{2}+1 \right)\\&& \\
     1&0 &\frac{n_l}{2}\left( \frac{n_l}{2}+1 \right)  \\&&\\
     1&1 &\frac{n_l}{2}\left( \frac{n_l}{2}+1 \right)  
\end{array}
\eea
The site index $(j)$ is omitted in the above equation as it is on one particular site. 
Within the average electric field ansatz, i.e for $n_l(j)=n_l\Rightarrow h_E(j)=h_E$ for all sites $j$, resulting,
\bea
H_E^{(\mathrm{approx})}=\frac{g^2a}{2}N h^0_E
\eea
where, $N$ is total number of staggered sites on the lattice and $h^0_E=\frac{n_l}{2}\left( \frac{n_l}{2}+1 \right)$. Note that, for $n_l\gg0$, one can actually consider $h^0_E= h_E\equiv \frac{n_l^2}{4} $. 

At any site $j$, the onsite electric energy $h_E(j)$ differ from $h^0_E$ iff $n_i(j)=0,n_o(j)=1$, and that difference, that is relevant in strong coupling regime (for $n_l>0$)  is given by:
\bea 
\Delta h_E= \frac{n_l+1}{2}\left( \frac{n_l+1}{2}+1 \right)-h^0_E= \frac{n_l}{2}+\frac{3}{4}.
\eea
This correction term to $H_E^{(\mathrm{approx})}$ is particularly important for strong as well as intermediate coupling regime, where we consider mean value of gauge flux, that is not very large compared to that considered in weak coupling regime. Within the mean field ansatz the total electric part of the LSH Hamiltonian Hamiltonian is given by:
\bea
H_E^{(\mathrm{LSH})}=\frac{g^2a}{2}\left[ N h^0_E + \sum_{\{j'\}}\left( \frac{n_l}{2}+\frac{3}{4} \right)\right] 
\eea
where, $\{j'\}$ denotes the sites with fermionic configuration $n_i(j')=0,n_o(j')=1$. 
In the bulk limit of the lattice, the occurrence of $j'$ will be $N/4$ for N site lattice. Hence, the total mean field electric Hamiltonian in the bulk limit is given by:
\bea
\label{HEmf}
H_E^{(\mathrm{mLSH})}=\frac{g^2a}{2}\left[ N\frac{n_l}{2}\left( \frac{n_l}{2}+1 \right)  + \frac{N}{4}\left( \frac{n_l}{2}+\frac{3}{4} \right)\right] ~~~~~
\eea

\textbf{Mass Hamiltonian:} The mass term (\ref{HMLSH}), being independent of gauge field configuration remain the same in the mean field ansatz, also for both the strong and weak coupling regime.
\bea
H_M^{(\mathrm{approx})}=m\sum_j (-1)^j(\hat n_i(j)+\hat n_o(j))
\eea

\textbf{Interaction Hamiltonian:} The matter-gauge field interaction term is the most complicated within LSH framework as detailed in (\ref{HILSH}). In the strong coupling limit of the theory, this particular term gives small contribution to the Hamiltonian (see subsection \ref{Sec:scalingHamiltonian}) and can be treated perturbatively. However, in the weak coupling regime, this term becomes significant. The purpose of the present approximation scheme is to bring the interaction Hamiltonian into simple form, yet describing matter gauge dynamics in the weak coupling regime. 

The approximation scheme that we follow is replacing the local loop quantum numbers $n_l(j)$ by a constant $n_l\gg 0$ at all of the lattice sites. The interaction Hamiltonian given in (\ref{HILSH}) can be written as,
\bea
H_I^{\mathrm{LSH}}=\frac{1}{2a}\sum_{j=0}^{N-2} h_I(j,j+1)
\eea
where, 
\bea
h_I(j,j+1)&=& h_I^1(j,j+1)+h_I^2(j,j+1)\nonumber \\ 
&&+h_I^3(j,j+1)+h_I^4(j,j+1)
\eea
Each of these terms, can be further decoupled into left $(L)$ and right $(R)$ parts located at site $j$ and site $j+1$ respectively,
\bea h_I^{[s]}(j,j+1)= h_I^{[s]}(L) h_I^{[s]}(R)~~, [s]=1,2,3,4.\eea
Now, considering each term separately, one would obtain the following:
\begin{widetext}

\bea
h_I^{[1]}(L)&=& \frac{1}{\sqrt{\hat n_l+\hat n_o(j)(1-\hat n_i(j))+1}}\hat \chi_o^+ (\lambda^+)^{\hat n_i(j)}\sqrt{\hat n_l+2-\hat n_i(j)}= \hat \chi_o^+ (\lambda^+)^{\hat n_i(j)} \hat C_1(L)  \\ 
h_I^{[2]}(L)&=& \frac{1}{\sqrt{\hat n_l+\hat n_o(j)(1-\hat n_i(j))+1}} \hat \chi_o^- (\lambda^-)^{\hat n_i(j)}\sqrt{\hat n_l+2(1-\hat n_i(j))} = \hat \chi_o^- (\lambda^-)^{\hat n_i(j)} \hat C_2(L) \\
h_I^{[3]}(L)&=& \frac{1}{\sqrt{\hat n_l+\hat n_o(j)(1-\hat n_i(j))+1}} \hat \chi_i^+ (\lambda^-)^{1-\hat n_o(j)}\sqrt{\hat n_l+2\hat n_o(j)}=\hat \chi_i^+ (\lambda^-)^{1-\hat n_o(j)} \hat C_3(L)\\
h_I^{[4]}(L)&=& \frac{1}{\sqrt{\hat n_l+\hat n_o(j)(1-\hat n_i(j))+1}} \hat \chi_i^- (\lambda^+)^{1-\hat n_o(j)}\sqrt{\hat n_l+1+\hat n_o(j))} =\hat \chi_i^- (\lambda^+)^{1-\hat n_o(j)} \hat C_4 (L)
\eea
and 
\bea
h_I^{[1]}(R)&=& \hat \chi_o^- (\lambda^+)^{1-\hat n_i(j+1)}\frac{\sqrt{\hat n_l+1+\hat n_i(j+1))}}{\sqrt{ \hat n_l+\hat n_i(j+1)(1-\hat n_o(j+1))+1}} = \hat \chi_o^- (\lambda^+)^{1-\hat n_i(j+1)} \hat C_1(R)\\
h_I^{[2]}(R)&=& \hat \chi_o^+ (\lambda^-)^{1-\hat n_i(j+1)} \frac{\sqrt{\hat n_l+2\hat n_i}}{\sqrt{ \hat n_l+\hat n_i(j+1)(1-\hat n_o(j+1))+1}}=\hat \chi_o^+ (\lambda^-)^{1-\hat n_i(j+1)}\hat C_2(R)\\
h_I^{[3]}(R)&=& \hat \chi_i^- (\lambda^-)^{\hat n_o(j+1)} \frac{\sqrt{\hat n_l+2(1-\hat n_o(j+1))}}{\sqrt{ \hat n_l+\hat n_i(j+1)(1-\hat n_o(j+1))+1}}=\hat \chi_i^- (\lambda^-)^{\hat n_o(j+1)}\hat C_3(R) \\
h_I^{[4]}(R)&=& \hat \chi_i^+ (\lambda^+)^{\hat n_o(j+1)}\frac{\sqrt{\hat n_l+2-\hat n_o(j+1)}}{\sqrt{ \hat n_l+\hat n_i(j+1)(1-\hat n_o(j+1))+1}} =\hat \chi_i^+ (\lambda^+)^{\hat n_o(j+1)} \hat C_4(R)\eea
\end{widetext}
The only approximation made in the above set of equations is $n_l(j),n_l(j+1)\rightarrow n_l$, where $n_l$ is the mean field value. The explicit operator form of the coefficients $\hat C_{[s]}(L/R)$'s are the following:
\begin{widetext}
\bea
\label{hE}
\begin{array}{|c|c|c|c|c|c|c|c|c|c|}
    \hline
    n_i & n_o&\hat C_{1}(L) &\hat C_{2}(L) &\hat C_{3}(L) &\hat C_{4}(L) &\hat C_{1}(R) &\hat C_{2}(R) &\hat C_{3}(R) &\hat C_{4}(R)  \\
    \hline
    &&&&&&&&&\\
     0&0 & 1 & 1& 1& \sqrt{\dfrac{n_l+1}{n_l+2}}& 1& \sqrt{\dfrac{n_l}{n_l+1}}& \sqrt{\dfrac{n_l+2}{n_l+1}}& \sqrt{\dfrac{n_l+2}{n_l+1}} \\
     &&&&&&&&&\\
     \hline
     &&&&&&&&& \\
       0&1 &\sqrt{\dfrac{n_l+2}{n_l+1}} & \sqrt{\dfrac{n_l+2}{n_l+1}}& \sqrt{\dfrac{n_l+2}{n_l+1}}& \sqrt{\dfrac{n_l+2}{n_l+1}}& 1& \sqrt{\dfrac{n_l}{n_l+1}}& \sqrt{\dfrac{n_l}{n_l+1}}& 1 \\
       &&&&&&&&&\\
       \hline
       &&&&&&&&& \\
        1&0 &\sqrt{\dfrac{n_l+1}{n_l+2}} & 1& 1& \sqrt{\dfrac{n_l+1}{n_l+2}}& 1& 1& 1& \sqrt{\dfrac{n_l+1}{n_l+2}} \\
        &&&&&&&&&\\
       \hline
       &&&&&&&&& \\
         1&1 &\sqrt{\dfrac{n_l+1}{n_l+2}} & 1& 1& 1 & \sqrt{\dfrac{n_l+2}{n_l+1}}&\sqrt{\dfrac{n_l+2}{n_l+1}}& \sqrt{\dfrac{n_l}{n_l+1}}& 1    \\
         &&&&&&&&&\\
       \hline
\end{array}
\eea
\end{widetext}
It is clear from the above set of coefficients that in the limit $n_l\gg0$, all of the coefficients can be approximated to be equal to identity operators, that is their leading order contribution. One can expand the coefficients and add corrections order by order. However, for this work as the very first step we keep ourselves confined to the leading order contribution. 

In this regime we also approximate $\lambda^{\pm}$ as identity operator as per the approximation, $ n_l+1 \approx n_l$.
Hence, the approximated interaction Hamiltonian is given by,
\bea
H_I^{(\mathrm{approx})}&=&\frac{1}{2a}\sum_j  \Big[ \chi_o^{+}(j)\chi_o^{-}(j+1)
+ \chi_o^{-}(j)\chi_o^{+}(j+1) \nonumber \\ &&
+\chi_i^{+}(j)\chi_i^{-}(j+1)
+\chi_i^{-}(j)\chi_i^{+}(j+1)\Big]
\eea
\end{document}